\pdfoutput=1

\documentclass[sigconf,nonacm]{acmart}

\AtBeginDocument{%
  \providecommand\BibTeX{{%
    \normalfont B\kern-0.5em{\scshape i\kern-0.25em b}\kern-0.8em\TeX}}}

\copyrightyear{2022}
\acmYear{2022}
\setcopyright{acmlicensed}\acmConference[AFT '22]{4th ACM Conference on Advances in Financial Technologies}{September 19--21, 2022}{Cambridge, MA, USA}
\acmBooktitle{4th ACM Conference on Advances in Financial Technologies (AFT '22), September 19--21, 2022, Cambridge, MA, USA}
\acmPrice{15.00}
\acmDOI{10.1145/3558535.3559772}
\acmISBN{978-1-4503-9861-9/22/09}

\usepackage{subcaption}
\usepackage{tikz}
\usepackage{lipsum}
\usepackage{balance}
\usepackage[export]{adjustbox}
\begin{document}

\title{Risks and Returns of Uniswap V3 Liquidity Providers}

\author{Lioba Heimbach}
\affiliation{%
\institution{ETH Zürich}
  \country{Switzerland}}
\email{hlioba@ethz.ch}
\author{Eric Schertenleib}
\affiliation{%
\institution{ETH Zürich}
  \country{Switzerland}}
\email{ericsch@ethz.ch}

\author{Roger Wattenhofer}
\affiliation{%
\institution{ETH Zürich}
  \country{Switzerland}}
\email{wattenhofer@ethz.ch}


\begin{abstract}
Trade execution on \emph{Decentralized Exchanges (DEXes)} is automatic and does not require individual buy and sell orders to be matched. Instead, liquidity aggregated in pools from individual liquidity providers enables trading between cryptocurrencies. The largest DEX measured by trading volume, Uniswap V3, promises a DEX design optimized for capital efficiency. However, Uniswap V3 requires far more decisions from liquidity providers than previous DEX designs.
    
In this work, we develop a theoretical model to illustrate the choices faced by Uniswap V3 liquidity providers and their implications. Our model suggests that providing liquidity on Uniswap V3 is highly complex and requires many considerations from a user. Our supporting data analysis of the risks and returns of real Uniswap V3 liquidity providers underlines that liquidity providing in Uniswap V3 is incredibly complicated, and performances can vary wildly. While there are simple and profitable strategies for liquidity providers in liquidity pools characterized by negligible price volatilities, these strategies only yield modest returns. Instead, significant returns can only be obtained by accepting increased financial risks and at the cost of active management. Thus, providing liquidity has become a game reserved for sophisticated players with the introduction of Uniswap V3, where retail traders do not stand a chance.
\end{abstract}





\maketitle

\section{Introduction}
Since the inception of Bitcoin~\cite{nakamoto2008bitcoin} in 2008 and later Ethereum~\cite{wood2014ethereum} in 2014, the capabilities of blockchains have significantly evolved. Most notably, the introduction of smart contracts by Ethereum enabled the blockchain to host an entire financial system commonly known as \emph{decentralized finance (DeFi)}. DeFi has shifted the widespread perception of cryptocurrencies being a tool for price speculation to a technology with the potential to revolutionize the future of finance. 

\emph{Decentralized exchanges (DEXes)} represent a DeFi cornerstone technology. DEXes allow users to exchange cryptocurrencies without giving up custody of their assets. Users interact directly with the smart contracts that build the DEXes. Trading is enabled by liquidity for each pair of tradeable cryptocurrencies reserved in a respective smart contract referred to as a \emph{liquidity pool}. Thus, when a user wishes to exchange $X$-tokens for $Y$-tokens, the user interacts with the respective liquidity pool, deposits $X$-tokens in the pool, and receives $Y$-tokens from the pool's liquidity. Individual liquidity providers supply the pool's liquidity by depositing both assets in the smart contract. For their service, the liquidity providers receive transaction fees from the trades supported by their liquidity. 

Most DEXes on the Ethereum platform implement a \emph{constant function market maker} (\emph{CFMM}) for automatic trade execution~\cite{2021uniswap,2021sushiswap,2021balancer,2021curve,angeris2020improved}. Uniswap~\cite{2021uniswap} is currently the biggest DEX on Ethereum~\cite{2022topdex} in terms of volume. There are two actively used versions of Uniswap: V2~\cite{adams2020uniswap} and V3~\cite{adams2021uniswap}. While Uniswap V2 implements a \emph{constant product market maker (CPMM)}, where the liquidity supplied by any liquidity provider supports trading on the entire price range, Uniswap V3 utilizes a new CPMM design intended to optimize capital efficiency. Uniswap V3 has acquired a significant market share throughout the past year and has overtaken its predecessor to become the DEX with the largest trading volume~\cite{2022coinmarketcap}. 

Liquidity providers on Uniswap V3 specify the price range in which they wish to supply their liquidity. Thus, the choices liquidity providers face have dramatically increased. In addition to choosing the pool to provide liquidity (as in Uniswap V2), in Uniswap V3, liquidity providers must also specify the position and width of the price range for which they wish to supply liquidity. This choice significantly impacts their expected returns, as well as the related financial risks\footnote{We focus on the financial risks stemming from price fluctuations and analyze them using historical data. Other risk, e.g., risks related to the protocol, are not considered.}. The increased complexity of providing liquidity begs the question of whether liquidity providing has become a game of sophisticated players or whether retail traders still stand a chance in comparison. To phrase it more pointedly, are retail traders looking for high returns as liquidity providers running a risk of losing it all?

In this work, we derive an analytical expression for the \emph{impermanent loss} of a liquidity position. Impermanent loss is a liquidity provider's risk of a decrease in the value of their liquidity position in comparison to the value of the initial assets. We find that the impermanent loss increases at a faster pace with concentrated liquidity. Furthermore, we present a theoretical model illustrating the complexities liquidity providers face. 

In addition, we analyze the performance of liquidity positions in the largest Uniswap V3 pools and show that the returns and risks of liquidity positions vary wildly. Both our model and the data highlight that due to the complexity of Uniswap V3, providing liquidity in price volatile pools requires both active management and high sophistication. Retail traders, unwilling to risk significant losses, should stick to simple strategies offering only low returns but also promising only small financial risks.

We further find that the interests of high profit seeking liquidity providers and the protocol may be misaligned. While the protocol seeks long term liquidity providers in order to offer traders sufficient market depth at any time, liquidity providers maximize their profit by actively managing their positions. The later could lead to a significant draining of liquidity in turbulent market situations.

\section{Related Work}
Several studies of the risks and returns of liquidity providers in the original CPMM exist. Evans~\cite{evans2020liquidity} studies the returns of liquidity providers and shows that they are capable of replicating various trading strategies and financial derivatives. In a further study, Evans et al.~\cite{evans2021optimal} find the optimal transaction fee for liquidity pools to attract liquidity. On the other hand, we empirically analyze the risks and returns of individual liquidity providers and illustrate the complexity facing liquidity providers in the novel CPMM design utilized by Uniswap V3. 

Heimbach et al.~\cite{Heimbach2021behavior} present an empirical study of the behavior of Uniswap V2 liquidity providers, as well as general risk and return metrics of Uniswap V2 pools. Their work shows the performance of liquidity providers is largely driven by the pool's price volatility. In contrast, our work is focused on the risk and returns of Uniswap V3 liquidity providers. We find that providing liquidity in Uniswap V3 is significantly more complex, exemplified by the stark difference in liquidity position returns dependent on their strategies, even within the same pool. 

Neuder et al.~\cite{neuder2021strategic} study strategic liquidity provision in Uniswap V3 and evaluate three classes of strategies for liquidity providers. However, their analysis makes several strong assumptions and, for instance, does not take into account the losses liquidity providers can face due to changing asset prices -- the sole driver of liquidity provider losses. In a similar line of work, Fritsch~\cite{Fritsch2021concentrated} quantifies the performance of a set of liquidity provider strategies by simulating them with historical trading data. We model the considerations liquidity providers must face when choosing the liquidity position in our work. Further, we empirically analyze the risks and returns of real Uniswap V3 liquidity providers. 

In a series of blog posts, Lambert~\cite{Lambert2021Guide,Lambert2021Guide2} investigates several challenges of providing liquidity on Uniswap V3 and makes the link between a Uniswap V3 liquidity position to financial derivatives. The presented analysis makes several simplifications to obtain an analytical solution. However, some of these assumptions are not supported by the actual data. We, on the other hand, generalize Lambert's ideas in our model and highlight the complexities faced by liquidity providers. 

A recent report presented by Loesch et al.~\cite{loesch2021impermanent} makes a first attempt at analyzing the returns of real Uniswap V3 liquidity positions and concludes that around 50\% of liquidity positions are losing money. However, their report has several shortcomings. For instance, they only consider the position's lifetime as a contributing factor to the liquidity position's return in their analysis. In our work, we account for multiple factors that influence the returns of liquidity providers on Uniswap V3 both theoretically and empirically, finding that with appropriate considerations providing liquidity on Uniswap V3 can be profitable.

\section{Constant Product Market Maker}

Uniswap V3~\cite{adams2021uniswap} functions as an \emph{automated market maker (AMM)}, i.e., trading is automatic, and a predefined algorithm controls the cryptocurrency prices. More specifically, Uniswap V3, like its predecessors Uniswap V1 and V2~\cite{adams2020uniswap}, utilizes the most widely adopted AMM subclass: \emph{constant product market maker (CPMM)}.

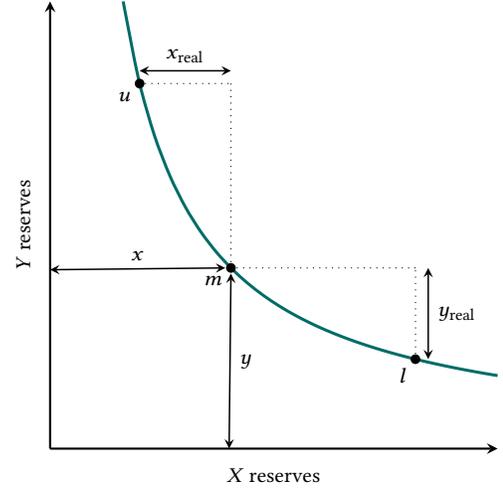
\begin{figure}[tbp]
  \centering
    \definecolor{color1}{HTML}{A01A7D}
\definecolor{color2}{HTML}{006C67}
\definecolor{color3}{HTML}{E55934}
\definecolor{color4}{HTML}{FFB100}  

  
  
  
  
  

\begin{tikzpicture}[scale=1.7]
  \draw[-stealth, thick] (0, 0) -- (3.5, 0) node[midway, below,sloped,inner sep = 7 pt] {\small $X$ reserves};
  \draw[-stealth, thick] (0, 0) -- (0, 3.5) node[midway, above,sloped,inner sep = 7 pt] {\small $Y$ reserves};
  \draw[scale=1, line width=0.4mm, domain=0.57142:3.5, smooth, variable=\x, color2] plot ({\x}, {2/\x});
  

   \node at (2.857,0.7)[circle,fill,inner sep=1.3pt](A) {} ;
   \node at (2.857,1.41421)[](A1) {} ;
   
  \node at (2.857,0.7)[anchor = north east](F) {\small $l$} ;

   \node at (.7,2.857)[circle,fill,inner sep=1.3pt](B) {} ;
   \node at (1.41421,2.857)[](B1) {} ;
  
   \draw [stealth-stealth, line width=0.2mm] (.7,2.957) -- (1.41421,2.957) node[midway,above] {\small$x_{\text{real}}$}  ;
   \node at (0.7,2.857)[anchor = north east](E) {\small $u$} ;
  \node at (1.41421,1.41421)[circle,fill,inner sep=1.3pt](C) {} ;

  \node at (1.41421,1.41421)[anchor = north east](G) {\small $m$} ;
  \draw [dotted] (A) -- (A1.center);
  \draw [dotted] (A1.center) -- (C);
  \draw [dotted] (B) -- (B1.center);
  \draw [dotted] (B1.center) -- (C);

  \draw [stealth-stealth, line width=0.2mm] (2.957,0.7) -- (2.957,1.41421) node[midway,right] {\small$y_{\text{real}}$}  ;
  
  
  \draw [stealth-stealth, line width=0.2mm] (0,1.4) -- (1.37,1.41421) node[midway,above] {\small$x$}  ;
  
  \draw [stealth-stealth, line width=0.2mm] (1.4,0) -- (1.41421,1.37) node[midway,right] {\small$y$}  ;
  
\end{tikzpicture}\vspace{0pt}
    
    \caption{CPMM price curve with virtual liquidity. The blue price curve ensures the constant product $x\cdot y$. The current marginal market price $S$ is at point $m$ and is given by the ratio of the virtual reserves, $S=y/x$.} \label{fig:curve}\vspace{0pt}
 \end{figure}

Uniswap aggregates liquidity in what is known as a \emph{liquidity pool} for each tradeable cryptocurrency pair. We note that Uniswap V2 and V3 allow for the creation of liquidity pools between any ERC-20 tokens --  an Ethereum standard for fungible tokens. Many individual liquidity providers supply the pool's liquidity, and their aggregated liquidity enables trading in the pool. The CPMM enforces that during trading, the product between the reserves of the pool's two cryptocurrencies stays constant, i.e., the pool's state moves along the price curve drawn in Figure~\ref{fig:curve}.

\begin{figure}[t]
\centering
  
  \begin{subfigure}{0.49\linewidth}
  
  \centering
   \definecolor{color1}{HTML}{A01A7D}
\definecolor{color2}{HTML}{006C67}
\definecolor{color3}{HTML}{E55934}
\definecolor{color4}{HTML}{FFB100}   
\begin{tikzpicture}[scale=0.2]
\filldraw[fill=color1!20!white, draw=color1,line width=0.4mm] (0,0) rectangle (14,4);
  \draw[-stealth, thick] (0, 0) -- (15, 0) node[midway, below,sloped,inner sep = 7pt] {\small price};
  \draw[-stealth, thick] (0, 0) -- (0, 8) node[midway, above,sloped,label distance =1.2cm] {\small liquidity};
	\node at (0,-0.9) {\small $0$};
     \node at (14,-0.9) {\small $\infty$};
\end{tikzpicture}\vspace{-4pt}
    \caption{Uniswap V2 liquidity} \label{fig:liquidityv2}
  \end{subfigure}%
  \hfill
  \begin{subfigure}{0.49\linewidth}
  \centering
    \definecolor{color1}{HTML}{A01A7D}
\definecolor{color2}{HTML}{006C67}
\definecolor{color3}{HTML}{E55934}
\definecolor{color4}{HTML}{FFB100}  

\begin{tikzpicture}[scale = 0.2]
\filldraw[fill=color1!20!white, draw=color1,  line width=0.4mm] (4,0) rectangle (12,7);
  \draw[-stealth, thick] (0, 0) -- (16, 0) node[midway, below,sloped,inner sep = 7pt] {\small price};
  \draw[-stealth, thick] (0, 0) -- (0, 8) node[midway, above,sloped,label distance =1.2cm] {\small liquidity};
	\node at (4,-0.9) {\small $S_l$};
     \node at (12,-0.9) {\small $S_u$};
\end{tikzpicture}\vspace{-4pt}
    \caption{Uniswap V3 liquidity position} \label{fig:liquidityv3}
\end{subfigure}\vspace{3pt}

\begin{subfigure}{\linewidth}
  \centering
    \definecolor{color1}{HTML}{A01A7D}
\definecolor{color2}{HTML}{006C67}
\definecolor{color3}{HTML}{E55934}
\definecolor{color4}{HTML}{FFB100}   
\begin{tikzpicture}[scale=0.2]
  \draw[-stealth, thick] (-3, 0) -- (-3, 11) node[midway, above,sloped,label distance =1.2cm] {\small liquidity};
	\node at (-2,-0.9) {\small $0$};
     \node at (20,-0.9) {\small $\infty$};

\filldraw[fill=color1!20!white, draw=color1, very thick] (8,0) rectangle (10,10);
\filldraw[fill=color1!20!white, draw=color1, very thick] (10,0) rectangle (12,9);
\filldraw[fill=color1!20!white, draw=color1, very thick] (-2,0) rectangle (0,0.5);
\filldraw[fill=color1!20!white, draw=color1, very thick] (0,0) rectangle (2,3);

\filldraw[fill=color1!20!white, draw=color1, very thick] (2,0) rectangle (4,4);
\filldraw[fill=color1!20!white, draw=color1, very thick] (6,0) rectangle (8,8);

\filldraw[fill=color1!20!white, draw=color1, very thick] (4,0) rectangle (6,7);

\filldraw[fill=color1!20!white, draw=color1, very thick] (16,0) rectangle (18,3);
\filldraw[fill=color1!20!white, draw=color1, very thick] (18,0) rectangle (20,1);

\filldraw[fill=color1!20!white, draw=color1, very thick] (14,0) rectangle (16,6);

\filldraw[fill=color1!20!white, draw=color1, very thick] (12,0) rectangle (14,7);
  \draw[-stealth, thick] (-3, 0) -- (21, 0) node[midway, below,sloped,inner sep = 7pt] {\small price};

\end{tikzpicture}\vspace{-4pt}
    \caption{collection of Uniswap V3 liquidity positions} \label{fig:liquiditys}\vspace{-4pt}
\end{subfigure}
\caption{Schematic representation of liquidity allocation on CPMMs. The points $S_l$ and $S_u$ specify the lower and upper bounds of the liquidity position's price range, respectively. Unlike on Uniswap V2, liquidity is not distributed uniformly across the entire price range but is determined by the choices of the liquidity providers.} \label{fig:liquidity}\vspace{0pt}
\end{figure}
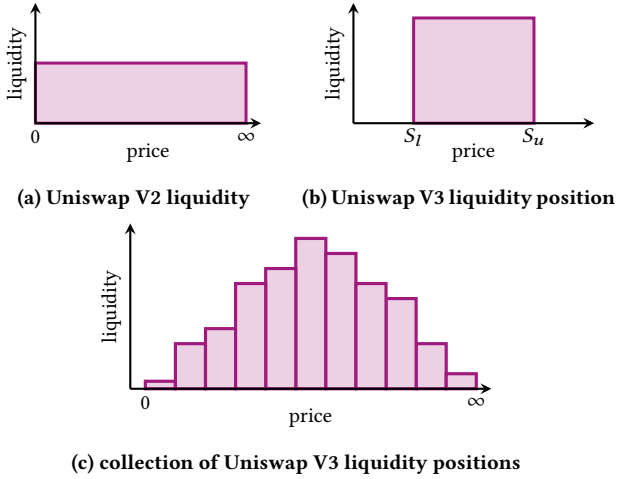

Uniswap V3's two predecessors, as well as SushiSwap~\cite{2021sushiswap}, employ the original CPMM design. To illustrate the pricing mechanism of the original CPMM, we consider a $X-Y$ pool between $X$-tokens and $Y$-tokens. If there are $x_{v2}$ $X$-tokens and $y_{v2}$ $Y$-tokens reserved in the pool, then the pool's \emph{marginal price} is given by $S=y_{v2}/x_{v2}$ and the pool's \emph{liquidity} is defined as $L_{v2}=\sqrt{x_{v2}\cdot y_{v2}}$~\cite{adams2020uniswap}. In this traditional implementation of CPMM, liquidity placed in the pool supports trading on the entire price range $(0,\infty)$ (cf. Figure~\ref{fig:liquidityv2}).

With Uniswap V3, Adams et al.~\cite{adams2021uniswap} introduce a novel CPMM design. Uniswap V3 liquidity providers specify the price range $[S_l,S_u]$ in which they wish to supply liquidity (cf. Figure~\ref{fig:liquidityv3}). Their liquidity then only supports trading within this price range. Thus, Uniswap V3 has an increased liquidity concentration around the current price and thereby increases the market's capital efficiency. 

We note that liquidity providers can only choose the price boundaries of their liquidity position from a predefined set of the pool's initialized \emph{ticks}. There is a tick at every integer exponent of $1.0001$, and the price of tick $i$ ($i \in \mathbb{Z}$) is given by 
$$S(i) = 1.0001^{i}.$$
Consequently, each tick is $0.01\%$ (one \emph{basis points (bps)}) away from each neighboring tick. Not every tick can be initialized in a pool, instead, only ticks with indexes divisible by the pool's predefined tick spacing ($t_s$) can be initialized. Figure~\ref{fig:liquiditys} shows a schematic representation of a liquidity allocation across a liquidity pool's price range. The liquidity is no longer constant across the entire price range, instead, only between the pool's initialized ticks. 

Uniswap V3 utilizes the concept of \emph{virtual reserves}, adjusted larger reserves, to describe the pool's behavior between two adjacent ticks $T_l$ and $T_u$. The virtual reserves act as if the liquidity in the entire pool matches that of the current price range. Thus, the virtual reserves of the current price range are a transformation of the range's real reserves that allow for the application of the constant product formula. In the following we consider a pool with reserves $x_{\text{real}}$ and $y_{\text{real}}$ between two adjacent ticks. The price range between ticks $T_l$ and $T_u$ is given by $[S_l,S_u]$. Instead of ensuring that the product between the reserves stays constant, the protocol ensures that the product of the virtual reserves $x$ and $y$ stays constant, i.e.,
$$x\cdot y=L^2,$$
where $L$ the \emph{liquidity} reserved in between ticks $T_l$ and $T_u$~\cite{adams2021uniswap}. Further, the \emph{marginal price} is given by $S= y/x$. The following relationship then holds between the virtual reserves, liquidity, and marginal price: 
$$x = \frac{L}{\sqrt{S}} \qquad y = L \sqrt{S}.$$ 
Thus, the virtual reserves behave according to the constant product price curve as shown in Figure~\ref{fig:curve}. Similarly, the real reserves can be obtained as follows: 
$$x_{\text{real}} =  x- \frac{L}{\sqrt{S_u}} =  \frac{L}{\sqrt{S}} - \frac{L}{\sqrt{S_u}} \qquad y_{\text{real}}=y-L \sqrt{S_l} =L \sqrt{S}-L \sqrt{S_l}.$$ 
Observe that the pool only needs to maintain sufficient reserves to support trading within the price boundaries $[S_l,S_u]$. Thus, the real reserves of $X$-token shrink as the value of $X$ in terms of $Y$ increases and are fully depleted at the upper price boundary $S_u$. The opposite holds for the real reserves of $Y$-token. 

We demonstrate Uniswap's trading mechanism by, again, considering a liquidity pool $X-Y$. The (virtual) reserves at the current price are given by $x$ $X$-tokens and $y$ $Y$-tokens. In Uniswap V2, a trader wishing to exchange $\delta_x$ $X$-tokens for $Y$-tokens,  will receive $\delta _y$ $Y$-tokens, where
\begin{equation*}
  \delta _{y}= y - \frac{x\cdot y}{x+(1-f)\delta_{x}} = \frac{y(1-f )\delta_{x}}{x+(1-f)\delta_{x}}. 
\end{equation*}
Here, $f$ is the transaction fee charged relative to the trader's input amount $\delta_x$~\cite{adams2020uniswap}. As long as the pool's price does not move across an initialized tick, the same holds in Uniswap V3. Otherwise, the trade executes at the available liquidity depth until it reaches the next initialized price tick. Then the remaining trade is completed at the available liquidity depth in the subsequent interval. This procedure re-applies until the entire trade input is swapped. 

The collected transaction fees are distributed pro-rata to the pool's liquidity providers, who have deposited liquidity at the price the asset pairs are trading at. We note that while the fees in Uniswap V2 were compounded, they are not in Uniswap V3. Further, observe that on Uniswap V3 the same asset pairs can be tradable in different pools that differ by the fee that is charged to traders. These are referred to as pool fee tiers. The possible tiers are $f\in \{0.01\%, 0.05\%,0.3\%,1\%\}$. Thus, the fees received by a liquidity provider in the current price range are not only dependent on the pool's volume and liquidity depth but also on its fee tier. Uniswap suggests the usage of lower transaction fees in pools with low relative price volatility between the two cryptocurrencies, such as two stablecoins. As suggested by the name, stablecoins are designed to have a stable price and are often pegged to the US\$. Higher transaction fees are suggested in pools with a high relative price volatility between the two cryptocurrencies.

\subsection{Liquidity Provision}

In the original CPMM design, liquidity providers only choose a pool and their liquidity supported trading on the entire price interval. In comparison to holding their assets, liquidity providers' returns were positively influenced by transaction fees and negatively by \textit{impermanent loss}. The latter describes the risk of a liquidity provider seeing the value of their reserved assets decrease in comparison to the value of the initial assets. In particular, price changes between the two reserved cryptocurrencies drive up a liquidity provider's impermanent loss. Thus, the returns of liquidity providers were influenced by a pool's volume and price volatility. 

In Uniswap V3, liquidity providers must also specify the range in which they wish to supply liquidity. It is, therefore, no longer only a pool's transaction fees and volatility but also the liquidity provider's price range that influences the returns. A liquidity provider's liquidity is only active when the pool's price $S$ is within her specified price range $[S_l,S_u]$. Whenever the liquidity is active, it is facilitating transactions and, in turn, earns fees from the pool's transactions. When, however, the price is outside the liquidity provider's price range, her liquidity is inactive. More specifically, a liquidity provider that places (virtual) liquidity $\tilde{L}$ into the pool $X-Y$ in the price range $[S_l,S_u]$ will have
\begin{equation}
    \tilde{x} _{\text{real}} = 
\begin{cases}
\tilde{L} \cdot \left( \frac{1}{\sqrt{S_l}}- \frac{1}{\sqrt{S_u}}\right)
 & S< S_l \\
\tilde{L}  \cdot \left( \frac{1}{\sqrt{S}}- \frac{1}{\sqrt{S_u}}\right) &S_l \leq S < S_u \\
0 & S\geq S_u
\end{cases}\label{eq:xreal}
\end{equation} 
$X$-tokens reserved in the pool when the pool's price is $S$, as well as 
\begin{equation}
\tilde{y}_{\text{real}}  = 
\begin{cases}
0 & S< S_l \\
\tilde{L} \cdot \left( \sqrt{S}- \sqrt{S_l}\right) &S_l \leq S < S_u \\
\tilde{L}  \cdot \left( \sqrt{S_u}- \sqrt{S_l}\right)& S\geq S_u
\end{cases}\label{eq:yreal}
\end{equation} 
$Y$-tokens~\cite{adams2021uniswap} \footnote{The virtual liquidity $\tilde{L}$ can be expressed in terms of the initial deposited value $V_0$ using $V_0 = S_0 \cdot x_0 + y_0$.}. Thus, when the price is within her price range, she holds both tokens in the pool, and her liquidity is active, i.e., she is earning fees. When, however, the price is outside the boundaries, she only holds one of the tokens. More specifically, when $S<S_l$ (the price of $X$ with respect to $Y$ decreases and is outside the boundary), her liquidity only consists of $X$-tokens. On the other side of her price boundary, the opposite holds true. Thus, both the location and width of the price range starkly influence a liquidity provider's return. While large price ranges decrease a liquidity provider's capital efficiency, the liquidity is likely active for longer periods. Small price ranges, on the other hand, increase a liquidity provider's capital efficiency, i.e., she earns more fees relative to her liquidity size when the price is in her range. At the same time, the price is generally more likely to fall out of her range more quickly. Thus, liquidity providers often readjust their liquidity positions following price changes. In the following, we provide a thorough analysis and evaluation of the risks and returns faced by Uniswap V3 liquidity providers.


\section{Analysis}
In this section, we discuss the factors influencing the performance of liquidity positions on Uniswap V3. Liquidity providers face several decisions when choosing their liquidity position. We provide a theoretical discussion of the implications of each of these choices. 

\subsection{Impermanent Loss}
We start by analyzing the main risk faced by liquidity providers: impermanent loss. The impermanent loss describes the loss in value of a liquidity position in comparison to holding the original assets as the price changes. We start by deriving the impermanent loss of a Uniswap V2 liquidity provider, which corresponds to a Uniswap V3 liquidity provider setting her price range to $(0,\infty)$. Consider a liquidity provider that places $\tilde{L}$ liquidity into a pool $X- Y$ when the pool's marginal price is $S_0$. Thus, the liquidity provider places $x_0 =  {\tilde{L}}/{\sqrt{S_0}}$ $X$-tokens and $y_0 =\tilde{L}\sqrt{S_0} $ $Y$-tokens in the pool. The value of the liquidity provider's position at a later point, when the pool's price is $S_1$, is given by: 
\begin{align*}
    V_{\text{v2,pos}}(\tilde{L}, S_1) &= S_1\cdot x_{1} + y_{1} = \frac{\tilde{L}}{\sqrt{S_1}}S_1 + \tilde{L}\sqrt{S_1} = 2 \tilde{L}\sqrt{S_1},
\end{align*}
where $x_1$ and $y_1$ are the liquidity provider's assets in the pool. The value of the liquidity provider's original assets, on the other hand, is given by
\begin{align*}
    V_{\text{v2,hold}}(\tilde{L},S_0, S_1) &= S_1\cdot x_{0} + y_{0} = \frac{\tilde{L}}{\sqrt{S_0}}S_1 +\tilde{L}\sqrt{S_0}.
\end{align*}

\begin{figure}[t]
\centering
\includegraphics[scale=1,right]{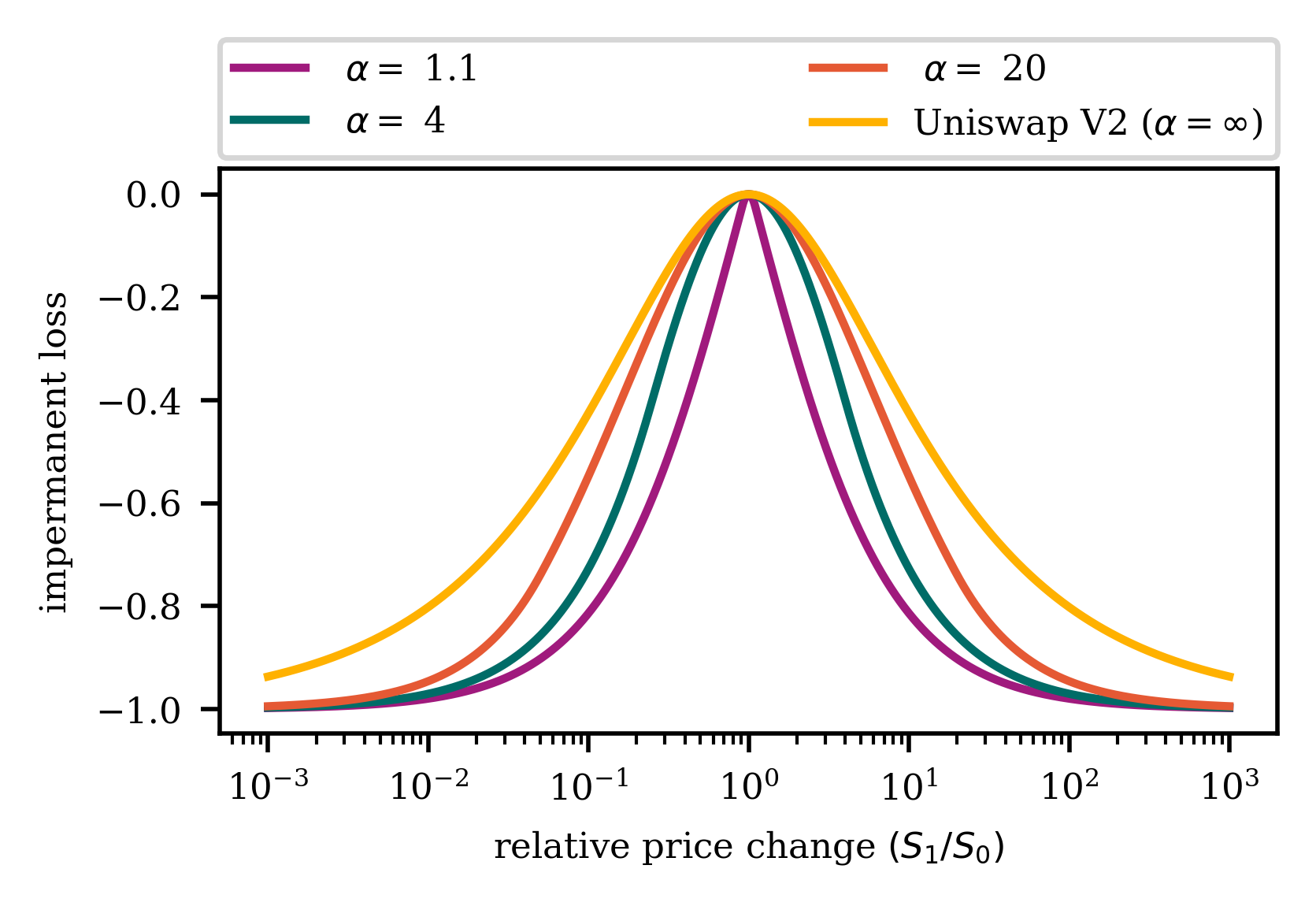}
\caption{Simulation of the impermanent loss of a liquidity position on Uniswap V2 and Uniswap V3 for different liquidity position widths. We set $S_l = 1/\alpha\cdot S_0$ and $S_u= \alpha\cdot S_0$.} \label{fig:iltheory}\vspace{0pt}
\end{figure}

Thus, we can obtain the impermanent loss as follows:
\begin{align*}
    \text{IL}_{\text{v2}} (S_0,S_1) &=   \frac{V_{\text{v2,pos}} -V_{\text{v2,hold}}}{V_{\text{v2,hold}}}\\
    &=\frac{2 \tilde{L}\sqrt{S_1}-\left(\frac{\tilde{L}}{\sqrt{S_0}}S_1 + \tilde{L}\sqrt{S_0} \right)}{\frac{\tilde{L}}{\sqrt{S_0}}S_1 + \tilde{L}\sqrt{S_0}}\\
    &= \left(\frac{2\cdot \sqrt{\frac{S_1}{S_0}}}{1+\frac{S_1}{S_0}}-1\right).
\end{align*}
We note that a liquidity providers impermanent loss is zero when $S_0=S_1$, i.e., the price is the same as at the initial time of liquidity injection. Otherwise, the impermanent loss is always negative\footnote{By convention the impermanent loss is smaller or equal to zero, thus, a non-zero impermanent loss is detrimental to the liquidity provider.}. We plot the impermanent loss as a function of the relative price change for a liquidity provider on the entire range, the equivalent of a Uniswap V2 liquidity provider, in yellow in Figure~\ref{fig:iltheory}.

In the following, we repeat the same steps to obtain the impermanent loss of a liquidity provider that supplies liquidity $L$ into the pool $X- Y$ in the price range $[S_l,S_u]$. The liquidity provider inserts the liquidity into the pool, when the pool's marginal price is $S_0$ and we, again, derive the impermanent loss at a later point in time, when the pool's marginal price is $S_1$. We start by obtaining the position value at price $S_1$ with the help of Equation~\ref{eq:xreal} and Equation~\ref{eq:yreal}:
\begin{align*}
    V_{\text{pos}}(\tilde{L}, S_1,S_l,S_u) &= S_1\cdot x_{1} + y_{1}\\
    &=\begin{cases}
        \tilde{L}\cdot S_1 \left( \frac{1}{\sqrt{S_l}}- \frac{1}{\sqrt{S_u}}\right)& S_1< S_l \\
        \tilde{L} \left(2  \sqrt{S_1}-\sqrt{S_l} - \frac{S_1}{\sqrt{S_u}}\right)& S_l \leq S_1 < S_u  \\
        \tilde{L} \cdot \left( \sqrt{S_u}- \sqrt{S_l}\right)& S_1\geq S_u.
    \end{cases}
\end{align*}
Here, $x_1$ $X$-tokens and $y_1$ $Y$-tokens are the liquidity position's real reserves in the pool. We obtain the value of the original inserted liquidity similarly:
\begin{align*}
    V_{\text{hold}}(\tilde{L}, S_0,S_1,S_l,S_u) &= S_1\cdot x_{0} + y_{0}\\
    &=\begin{cases}
        \tilde{L}\cdot S_1 \left( \frac{1}{\sqrt{S_l}}- \frac{1}{\sqrt{S_u}}\right)& S_0< S_l \\
        \tilde{L} \left(\frac{S_0+S_1}{\sqrt{S_0}}-\sqrt{S_l} - \frac{S_1}{\sqrt{S_u}}\right)& S_l \leq S_0 < S_u  \\
        \tilde{L} \cdot \left( \sqrt{S_u}- \sqrt{S_l}\right)& S_0\geq S_u,
    \end{cases}
\end{align*}
where $x_0$ $X$-tokens and $y_0$ $Y$-tokens are the reserves initially placed in the pool. The impermanent loss can then be obtained from the two preceding expression as follows:

$$IL(S_0, S_1,S_l,S_u)=   \frac{V_{\text{pos}} -V_{\text{hold}}}{V_{\text{hold}}}.$$

We simulate the impermanent loss of a liquidity provider with price range $[1/\alpha\cdot S_0,\alpha\cdot S_0]$ for $\alpha \in [1.1,4,20]$ in Figure~\ref{fig:iltheory}. Notice that the smaller $\alpha$, i.e., the tighter the price range, the faster the impermanent loss increases. Thus, not only do liquidity providers run a higher risk of their liquidity becoming idle, when the pool's price moves outside their interval, but their impermanent loss increases more quickly as well. The liquidity of liquidity providers with a wider price interval, on the other hand, is less capital efficient. They earn less fees for their liquidity, when the price is in the interval.

\begin{figure}[t]
\centering
\includegraphics[scale=1,right]{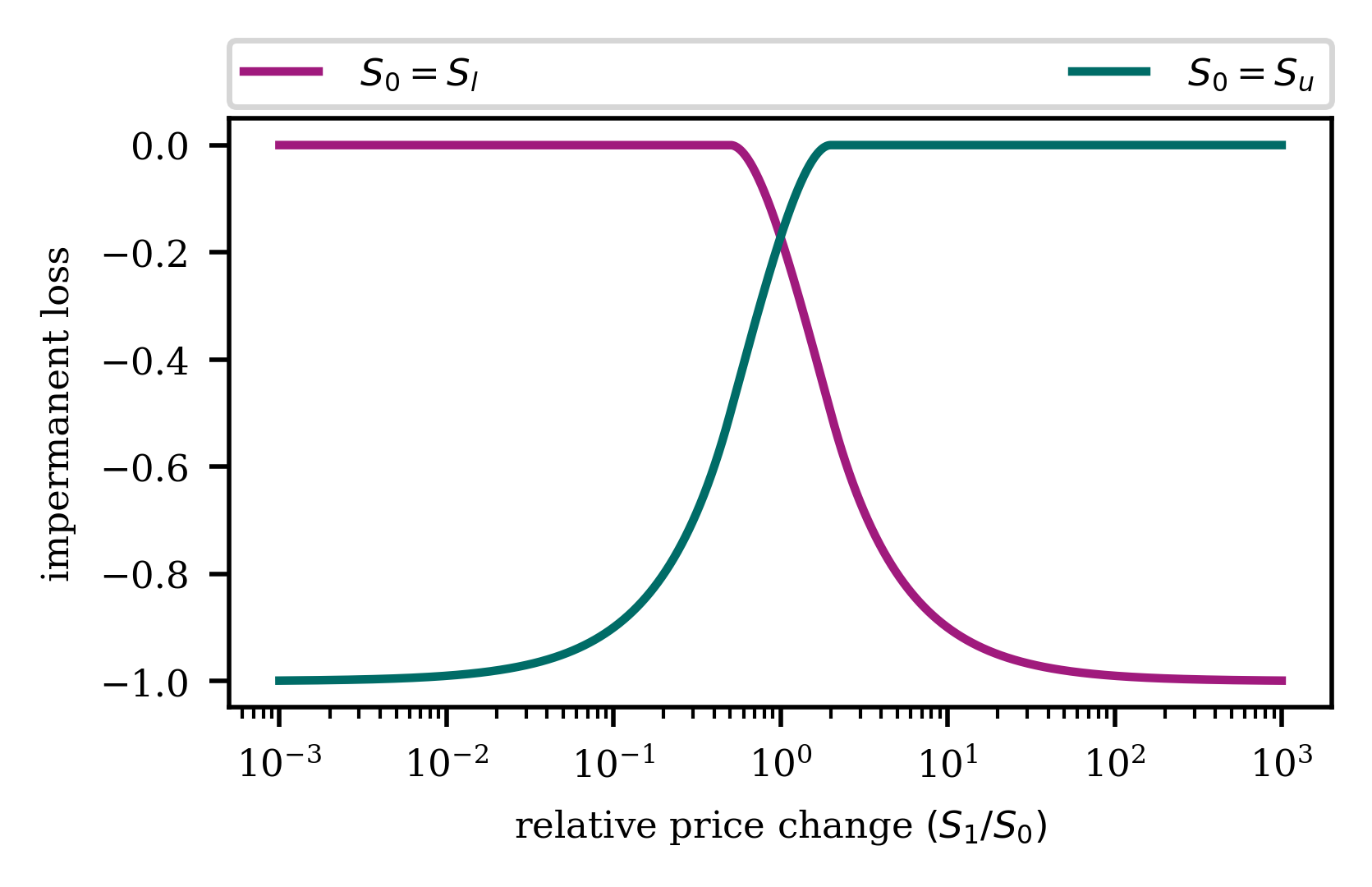}
\caption{Simulation of the impermanent loss of a liquidity position minted outside the current price on Uniswap V3. The respective price ranges are set $[S_0,2\cdot S_0]$ and $[0.5\cdot S_0, S_0]$.} \label{fig:iloutsiderange}\vspace{-0pt}
\end{figure}

Figure~\ref{fig:iltheory} shows the impermanent loss of a liquidity provider who enters the pool when the price is in the middle of her interval. In Figure~\ref{fig:iloutsiderange}, on the other hand, we show the impermanent loss of a liquidity provider that enters the pool, when the price is on the edge of her interval: $S_0 = S_l$ (drawn in green) and $S_0= S_u$ (drawn in violet). Notice that when the price remains outside the interval, i.e., $S_1 \leq S_l$ for $S_0 =S_l$, the impermanent loss is zero. However, unless the price moved in and out of the interval in the meantime, the liquidity provider also did not earn any fees. As soon as the price moves into the interval and beyond, the impermanent loss builds up quickly.

When choosing where to provide liquidity, Uniswap V3 liquidity providers must account for risks stemming from the impermanent loss, which are only negligible when little to no price movements are expected for the pool.

\subsection{Selection of Liquidity Position}\label{sec:return}

In addition to the negative influence of the experienced impermanent loss, a liquidity provider's return $R$ is positively influenced by the fees earned $F$ and given by 
\begin{equation}
R(S_0, S_1,S_l,S_u,F)=   \frac{V_{\text{pos}} +F-V_{\text{hold}}}{V_{\text{hold}}}.\label{eq:return}
\end{equation}

We note that the return compares the value of the initial assets deposited into the liquidity position to the value of the liquidity position (including the earned fees). This allows the performance comparison of owning the two tokens and providing liquidity vs. just owning the two tokens. This performance measurement is more suited than comparing it to a fixed currency, for example, US\$, as it is not dominated by the cryptocurrencies price evolution compared to this fixed currency, but rather allows to pinpoint the return that stems from the actual decision of providing liquidity.

The fees collected are distributed pro-rata to the pool's liquidity providers, who have deposited liquidity at the price the asset pairs are trading at. We note that on Uniswap the same asset pairs can be tradable in different pools that differ by the fee that is charged to traders. These are referred to as pool fee tiers. The possible tiers are $f\in \{0.01\%, 0.05\%,0.3\%,1\%\}$. Thus, the fees received by a liquidity provider in the current price range are not only dependent on the pool's volume and liquidity depth but also on its fee tier. 

Additionally, to garner fees, the selection of the liquidity position's price range is crucial. For pairs of stablecoins, the acquired fees are directly determined by the price range and trading volume, as the impermanent loss is insignificant. Thus, liquidity providers deposit in a narrow price range near the stable ratio. However, in volatile pools, like Ether in US\$, selecting a suitable liquidity position is more challenging as the price is more likely to exit the chosen price range leading to a higher impermanent loss and no further fees being collected.

\begin{figure}[t]
\centering
\includegraphics[scale=1,right]{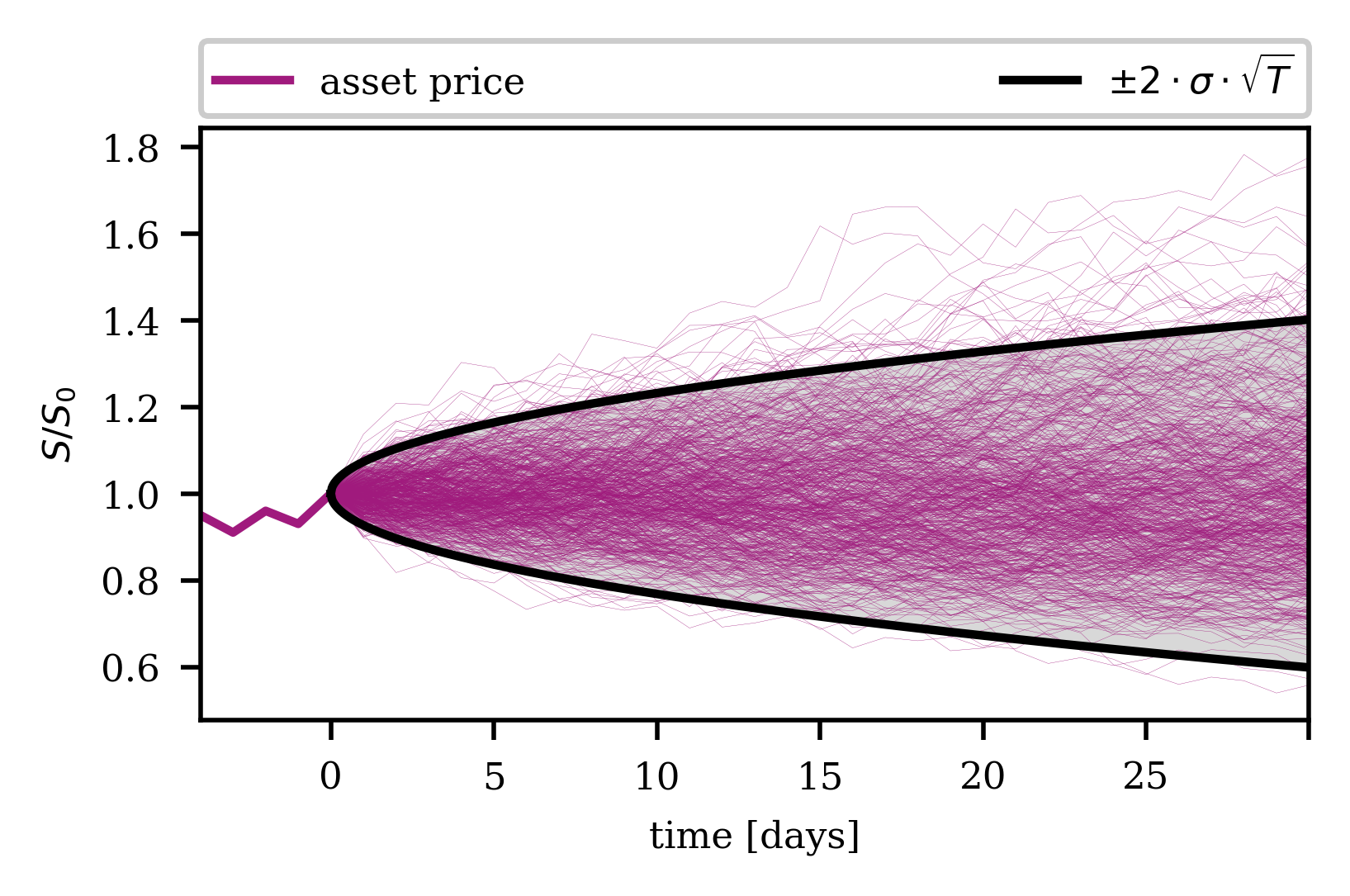}
\caption{Simulation of daily asset price paths over 30 days relative to initial price $S_0$. The shaded area shows 2-$\sigma$ level of the asset price distribution. Note that the width of this area grows with the square root of time (geometric Brownian motion). In this plot only 600 of 40'000 paths used in total are shown.} \label{fig:brown}\vspace{0pt}
\end{figure}

The feature that fees are only collected as long as the price remains within a certain bandwidth, necessitates the liquidity provider to gauge the probability thereof. This requirement naturally suggests adopting methods used for pricing financial derivatives where short- and medium-term price predictions are essential. The most well-known model for modeling the future development of the price of a risky asset $S(t)$ is the Black-Scholes market model, where the price is represented as an Itô process satisfying the stochastic differential equation
$$dS(t) = \mu S(T) dt + \sigma S(t) dW(t), $$
where $\mu$ is called the drift of the asset price and determines the expectation value of the future price, i.e., $\mathbb{E}[S(t)]=S_0 exp(\mu t)$~\cite{BlackScholes}. The volatility of the asset price is denoted as $\sigma$, and $W(t)$ is a Wiener process. Note that the term \emph{risky asset} is adopted from finance, but in our case, the choice of which coin in a pool is considered the risky asset is arbitrary. Rather, it just represents a choice of measuring the price of that coin $S(t)$ with respect to the other coin. Throughout this section, we pick an annual volatility of $70\%$, which is reasonable for an asset like Ether measured in US\$ (see for example~\cite{2022GVOL}). Furthermore, we choose the drift to be zero, as the short-term price movements are mainly governed by the volatility. Finally, we use the terminology from finance and say that a liquidity position is \textit{in the money (ITM)} if the asset price is within the price range the liquidity provider chose. In the opposite case, we say the position is \textit{out of the money (OTM)}.

Given the initial asset price $S(0)$ the above differential equation has the formal solution
\begin{equation}
S(t) = S(0) \exp \left(\mu t - \frac{\sigma^2}{2} t + \sigma W(t) \right).
\label{eq: Black-Scholes-solution}\vspace{-0pt}
\end{equation}

\begin{figure}[t]
\centering
\includegraphics[scale=1,right]{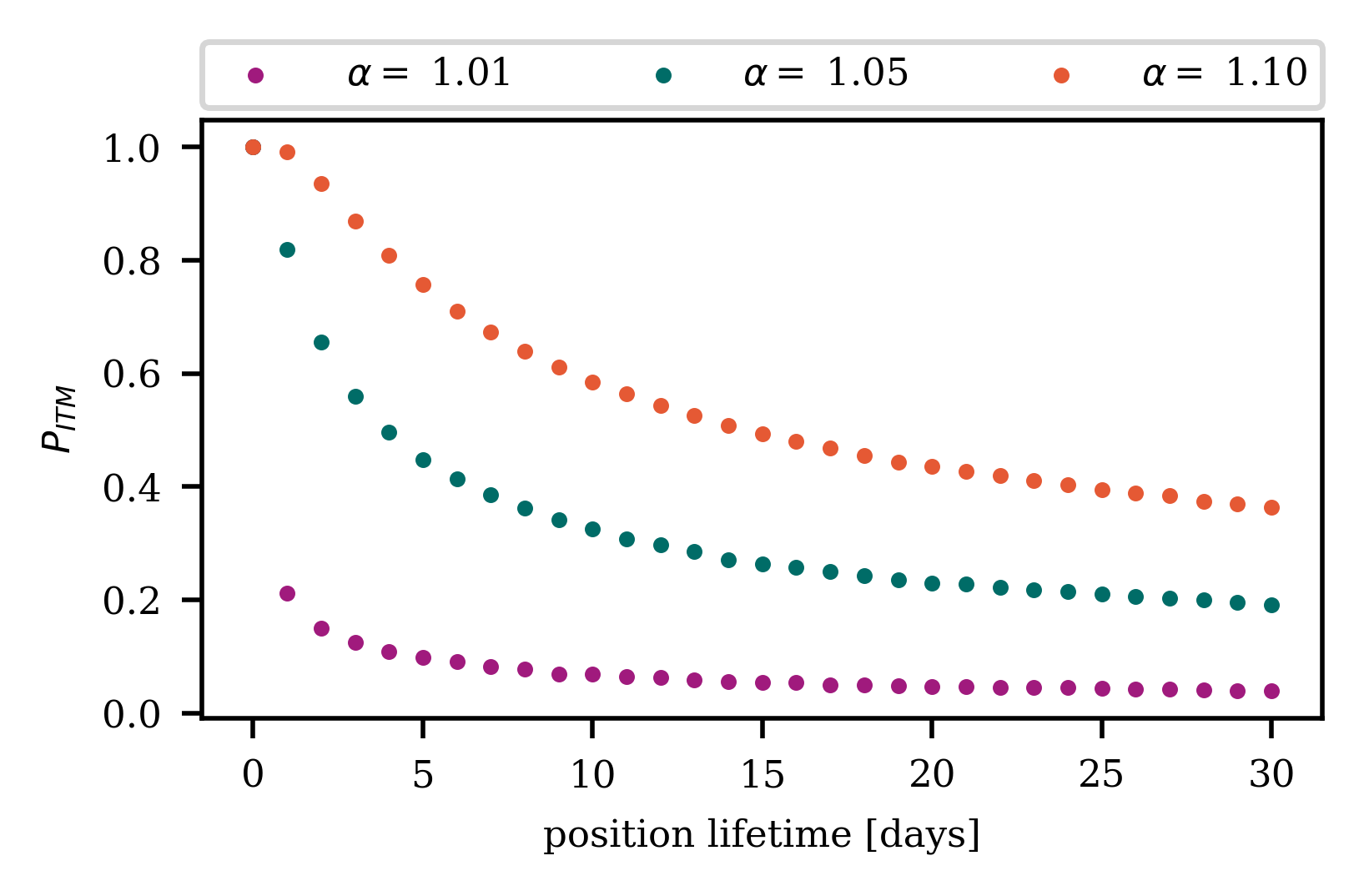}
\caption{Probability that the asset price remains within the liquidity position as a function of time for different $\alpha$. For a volatile asset the probability that the asset moves out of the selected range increases over time.}
\label{fig:PITMvst}
\end{figure}

As the liquidity provider only collects fees if the price remains within its selected price range, we are interested in a suitable selection of the liquidity range. To this end, we numerically calculate 40'000 possible future daily asset price paths using Equation~\ref{eq: Black-Scholes-solution} (cf. Figure~\ref{fig:brown}) and compute the probability that after time $T$ the asset has remained in the liquidity range determined by $\alpha$, where as in Figure~\ref{fig:iltheory}, we assume that the initial price $S_0$ is related to the lower and upper bound of the liquidity position by $S_l=S_0/\alpha$ and $S_u=\alpha S_0$, respectively. 

This probability $P_\mathrm{ITM}(t; \alpha)=P(S_0/\alpha<S(t)<\alpha S_0)$ is shown in Figure~\ref{fig:PITMvst} as a function of time for different $\alpha$. As expected, it becomes more likely that the position is OTM with passing time and a larger $\alpha$ means that the position is likely to remain ITM for longer.

\begin{figure}[t]
\centering
\includegraphics[scale=1,right]{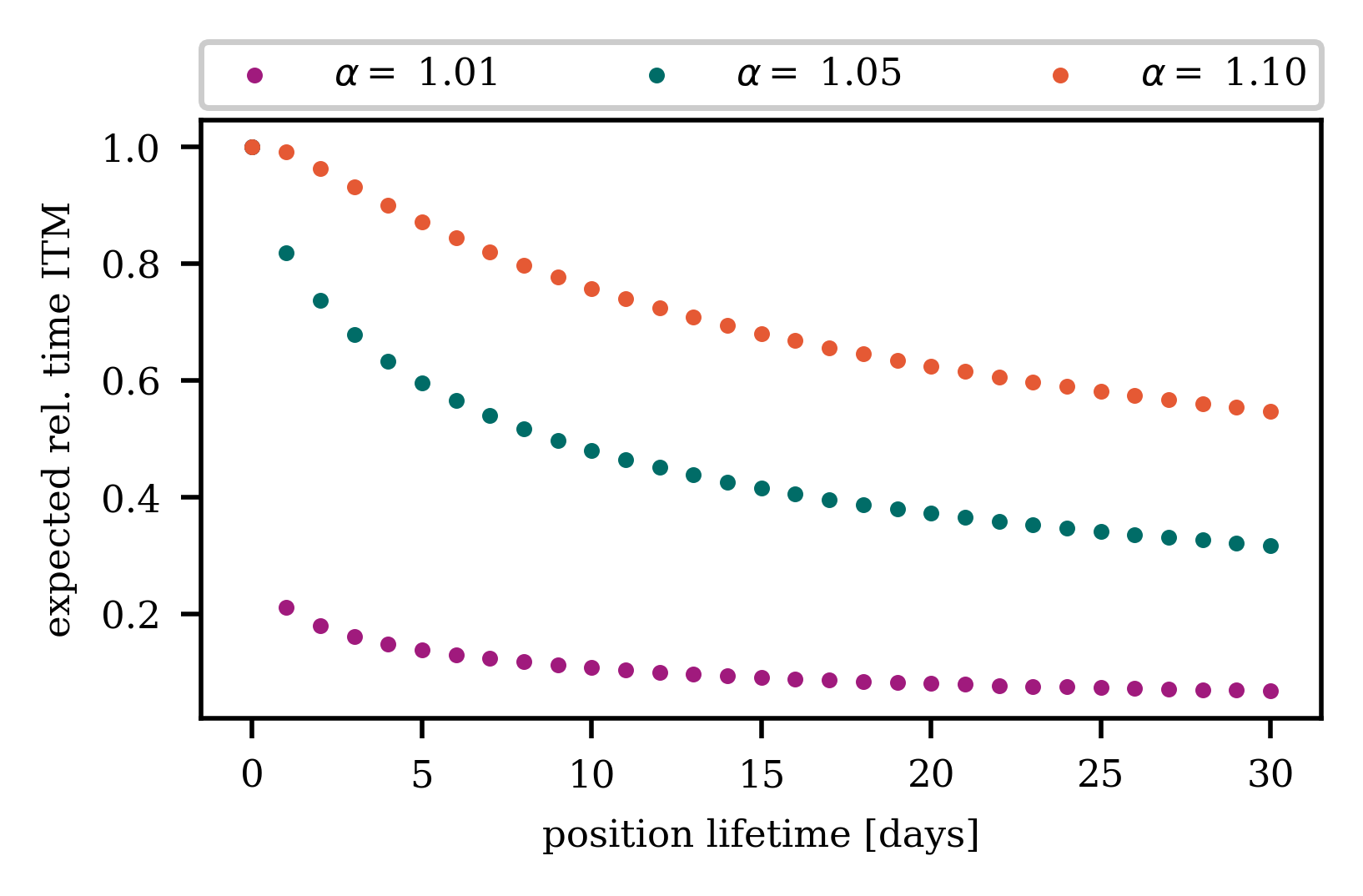}
\caption{Expected proportion of time the asset price lies within liquidity position for different $\alpha$. A wider position is expected to remain in the money, and hence collecting transaction fees, for a longer fraction of time.}
\label{fig:expectedtimeITM}\vspace{0pt}
\end{figure}

While the probability that a position remains ITM is important to gauge for a liquidity provider, she is more interested in the time the price spends ITM as, during this time, she profits from each trade that occurs in the pool. Given the probability that the price is ITM at time $t_i$, $P_\mathrm{ITM}(S_l<S(t_i)<S_u)$, we can compute the expected time the position is ITM relative to the total time passed $t_n$  
$$\mathbb{E}[T_\mathrm{ITM}(t_n; \alpha)] =  \frac{1}{t_n} \sum_{i=0}^{i=n} P_\mathrm{ITM}(t_i; \alpha) \Delta t,$$
where $\Delta t$ is the discrete time step (in our case one day). Figure~\ref{fig:expectedtimeITM} shows the fraction of time the price is in ITM as a function of time passed. Again, the relative expected time ITM decays over time and is smaller for narrower position

Figures~\ref{fig:PITMvst} and \ref{fig:expectedtimeITM} illustrate the relevance of both the time the liquidity provider keeps the position active as well as the width of the position's price range. While a large width reduces the probability that the position drops out of its price range, the fees collected per trade are adversely affected by a widely spread liquidity position. To illustrate this trade-off, we can consider the following simplified model. Assume that near the current price, the liquidity distribution of the pool is uniformly distributed, i.e., the same amount of liquidity in each tick. Then the fees collected by a position ITM are inversely proportional to $\alpha$. On the other hand, the fees collected are proportional to the trading volume multiplied by the time in the money $T_\mathrm{ITM}$. Thus, in this model, the total fee a liquidity provider collects is proportional to the time in the money divided by $\alpha$
$$F\propto T_\mathrm{ITM}/\alpha.$$

\begin{figure}[t]
\centering
\includegraphics[scale=1,right]{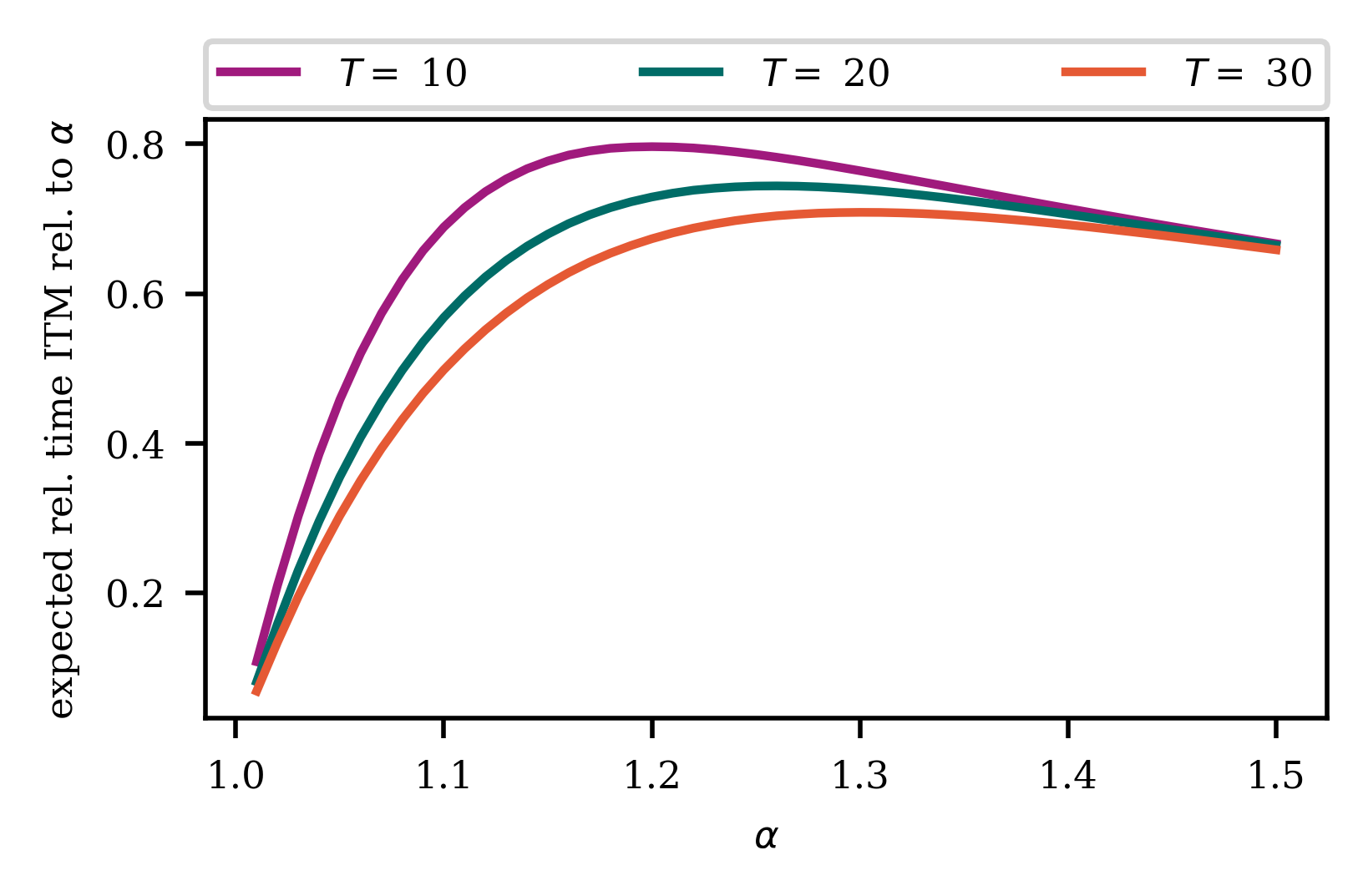}
\caption{Simulation of the proportion of time the asset price remains within the liquidity position relative to the liquidity width $\alpha$ after time $T$. As the collected fees grows with the time ITM and decays with position width the ratio of the time ITM relative to $\alpha$ is indicative for the collected fee. Thus, the longer a liquidity provider intends to keep her liquidity position in a volatile asset active, the larger the width should be chosen.}
\label{fig:TITMvsalpha}\vspace{-0pt}
\end{figure}

This quantity is depicted in Figure~\ref{fig:TITMvsalpha} as a function of $\alpha$ at different times $T$. The plot shows that there is an optimal $\alpha$, which depends on time, illustrating that the liquidity provider should carefully select their liquidity position width.

We conclude this section by noting that the liquidity providers in pools not consisting of two stablecoins face a complex problem. The liquidity provider should first have a prediction of future price and volatility developments of the selected pool. Therewith, she chooses the price range of her liquidity position, bearing in mind the time she intends to keep her funds locked, as well as the expected impermanent loss suffered. Additionally, she should also consider the liquidity distribution of the whole pool. Once the position is open, the liquidity provider must constantly monitor the position, update her predictions for the new market conditions and decide when it is best to withdraw the funds. This complexity illustrates that successfully providing liquidity in a volatile pool requires a high level of sophistication. 

\section{Real-World Measurements}
With the complexities faced by liquidity providers when selecting a liquidity position on Uniswap V3 in mind, we analyze the historic Uniswap V3 liquidity positions to understand the impact of these considerations. 

\subsection{Data Collection}

We analyze Uniswap V3 data to measure and understand the nature of the risks and rewards awaiting liquidity providers. The first Uniswap V3 pool was launched in block 12'369'739. Thus, we analyze the data beginning with block 12'369'739 (May 4, 2021) up until block 14'497'033 (last block on March 31, 2022). We collect data from the Ethereum blockchain by launching an erigon client. More specifically, we filter the event logs for all events related to Uniswap V3.

Our data analysis focuses on the four biggest cryptocurrencies on Uniswap V3 in terms of total value locked: WETH, WBTC, USDC, and DAI~\cite{2021uniswap}\footnote{Note that wrapped assets, e.g., WETH, by design, have the same value as the underlying, e.g., Ether. For our purpose, we can therefore consider them equivalent.}. Not only are the pools between these four cryptocurrencies amongst the largest in terms of volume and total value locked~\cite{2021uniswap}, but they also allow us to investigate varying patterns between different types of pools. 

With the introduction of Uniswap V3, currency pairs are classified as \emph{stable}, \emph{normal} or \emph{exotic} depending on the relative price volatility of the assets~\cite{2021Provide}. Stable pairs are characterized by little to no price changes between the pair's two cryptocurrencies, while we can expect significant price volatility between the two assets of a normal pair. Finally, for exotic pairs, at least one asset is not an established cryptocurrency, and the relative price between the two assets can fluctuate wildly. In their study of Uniswap V2 liquidity pools, Heimbach et al.~\cite{Heimbach2021behavior} highlight the differences in the returns and risks of liquidity providers depending on the pair's category, i.e., the pair's price volatility. These findings demonstrate the necessity of comparing the performance of liquidity providers based on the price volatility of the respective asset pair rather than over the entire protocol. Our selection of pools covers the largest pools both with respect to trading volume and total value locked. They make up approximately a third of the total liquidity on Uniswap. The performance of the liquidity position in these pools is thus indicative of the typical risks and returns faced by liquidity providers on Uniswap V3. We further note that our analysis is on the level of individual liquidity positions, as opposed to wallets or entities.

\subsection{General Liquidity Pool Statistics}\label{sec:generalstatistics}

We commence the data analysis by extracting general statistics of liquidity positions in the three Uniswap V3 pools with the highest total value locked at the time of writing~\cite{2021uniswap}: USDC-WETH (f= 0.3\%), WBTC-WETH (f= 0.3\%) and DAI-USDC (f= 0.01\%). USDC-WETH and WBTC-WETH are normal pools, as at least one of the pools' assets is subject to significant price movements. As opposed to exotic pools, all pool assets are established cryptocurrencies. DAI-USDC, on the other hand, is a stablecoin pool, as both the pool's tokens are pegged to the US\$ and, thus, only small price fluctuations are expected. The transaction fees levied by normal pools tend to be higher than the transaction fees charged by stable pools due to higher risks involved for liquidity providers stemming from the impermanent loss. We further note that for each of the three token pairs, multiple pools with different fee tiers exist that hold significant liquidity on Uniswap V3. We only plot the general liquidity position statistics for the pools with the highest liquidity for better visibility. In the later performance analysis (cf. Section~\ref{sec:performance}), we will include additional fee tiers for each token pair.  
 
\begin{figure}[t]
\centering
\begin{subfigure}[]{1\linewidth}
\includegraphics[scale=1,right]{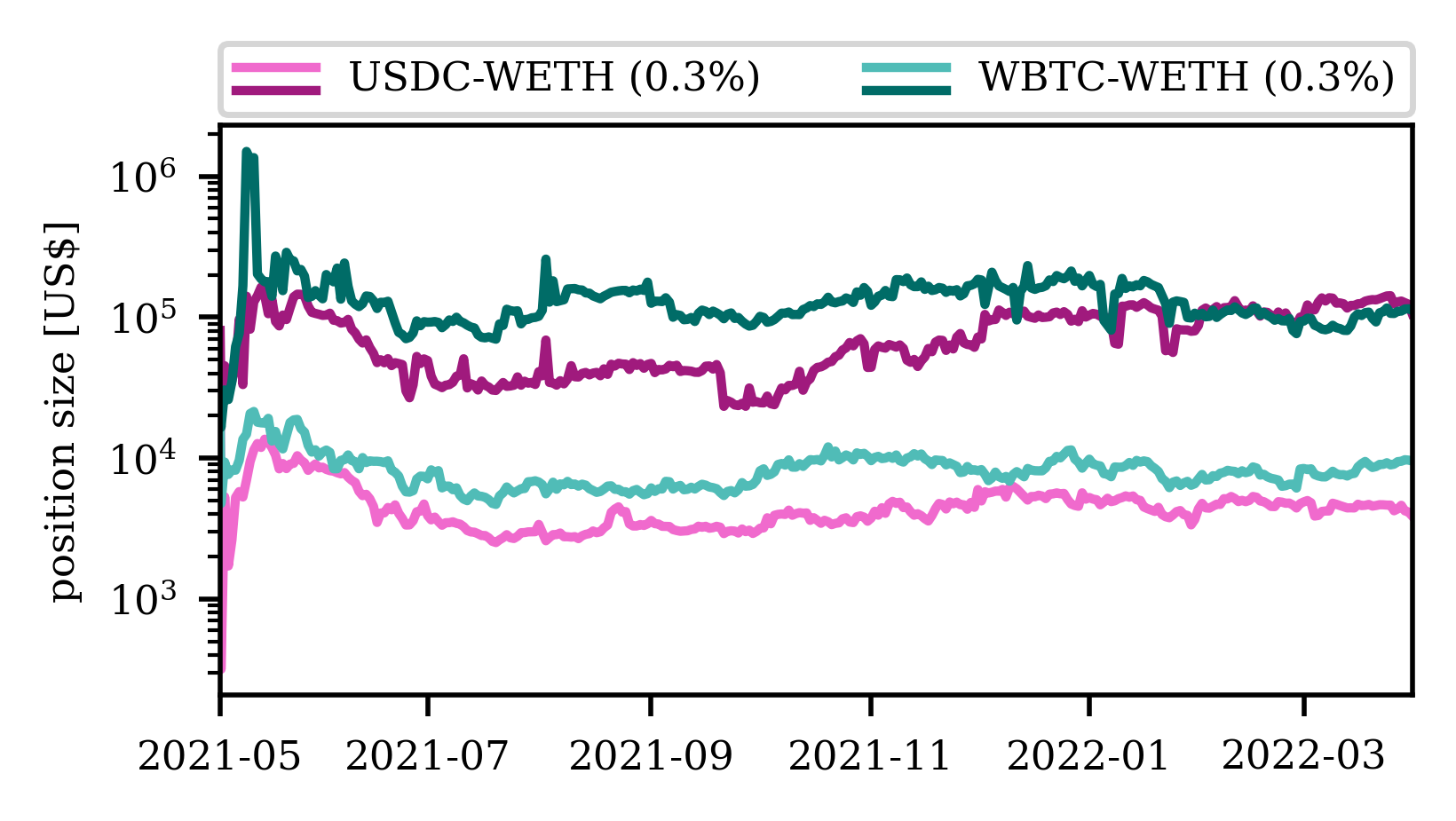}\vspace{-0pt}
\caption{USDC-WETH (f= 0.3\%) and WBTC-WETH (f= 0.3\%)} \label{fig:positionsizeETHBTC}
\end{subfigure}   
\begin{subfigure}[]{1\linewidth}
\includegraphics[scale=1,right]{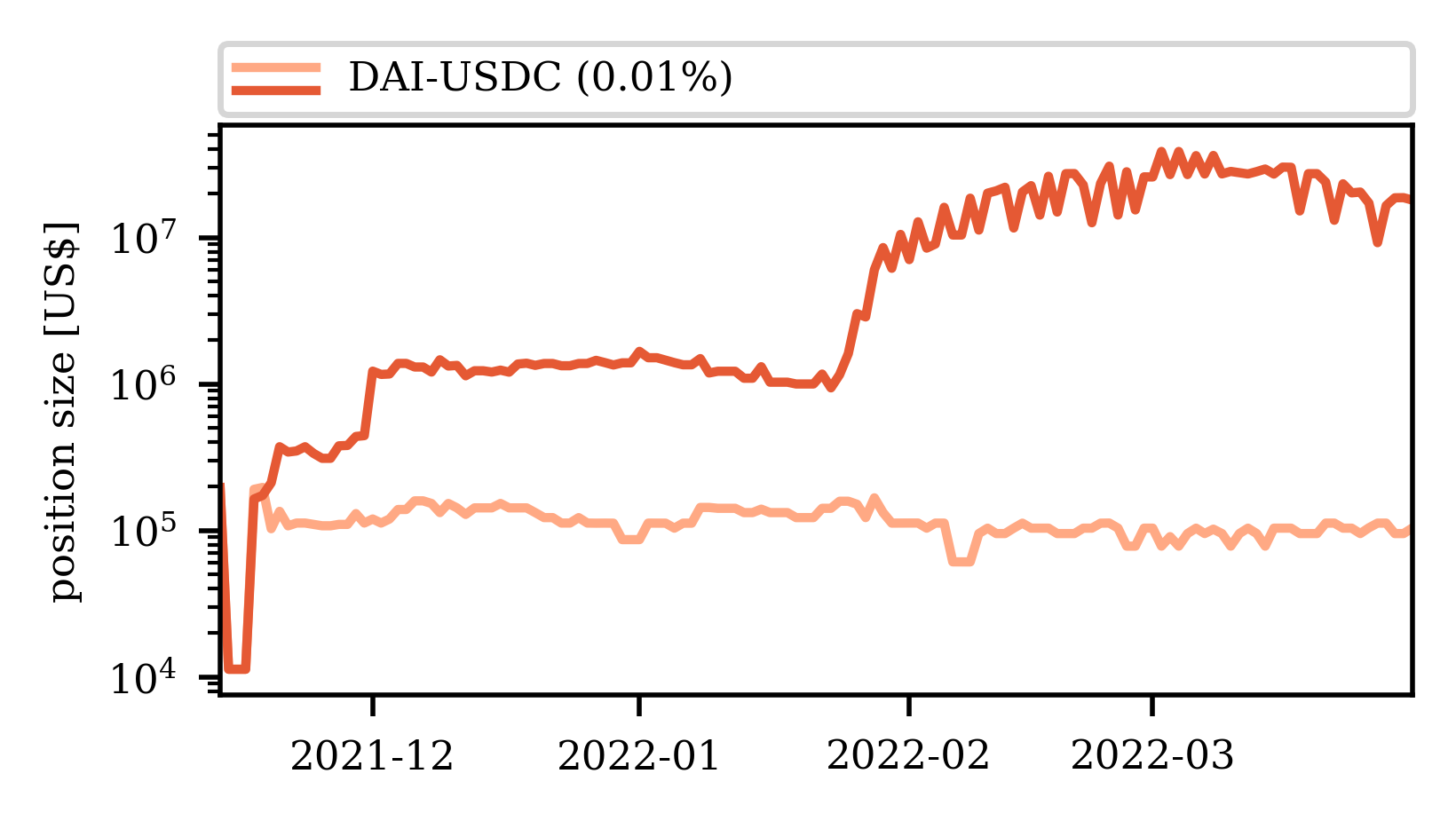}\vspace{-4pt}
\caption{DAI-USDC (f= 0.01\%)} \label{fig:positionsizestable}
\end{subfigure} 
\caption{Median (lighter lines) and mean (darker lines) position size over time in three Uniswap V3 pools. Observe the large difference (factor ten) between the median and mean.} \label{fig:positionsize}\vspace{-4pt}
\end{figure}

\begin{figure}[t]
\centering
\begin{subfigure}[]{1\linewidth}
\includegraphics[scale=1,right]{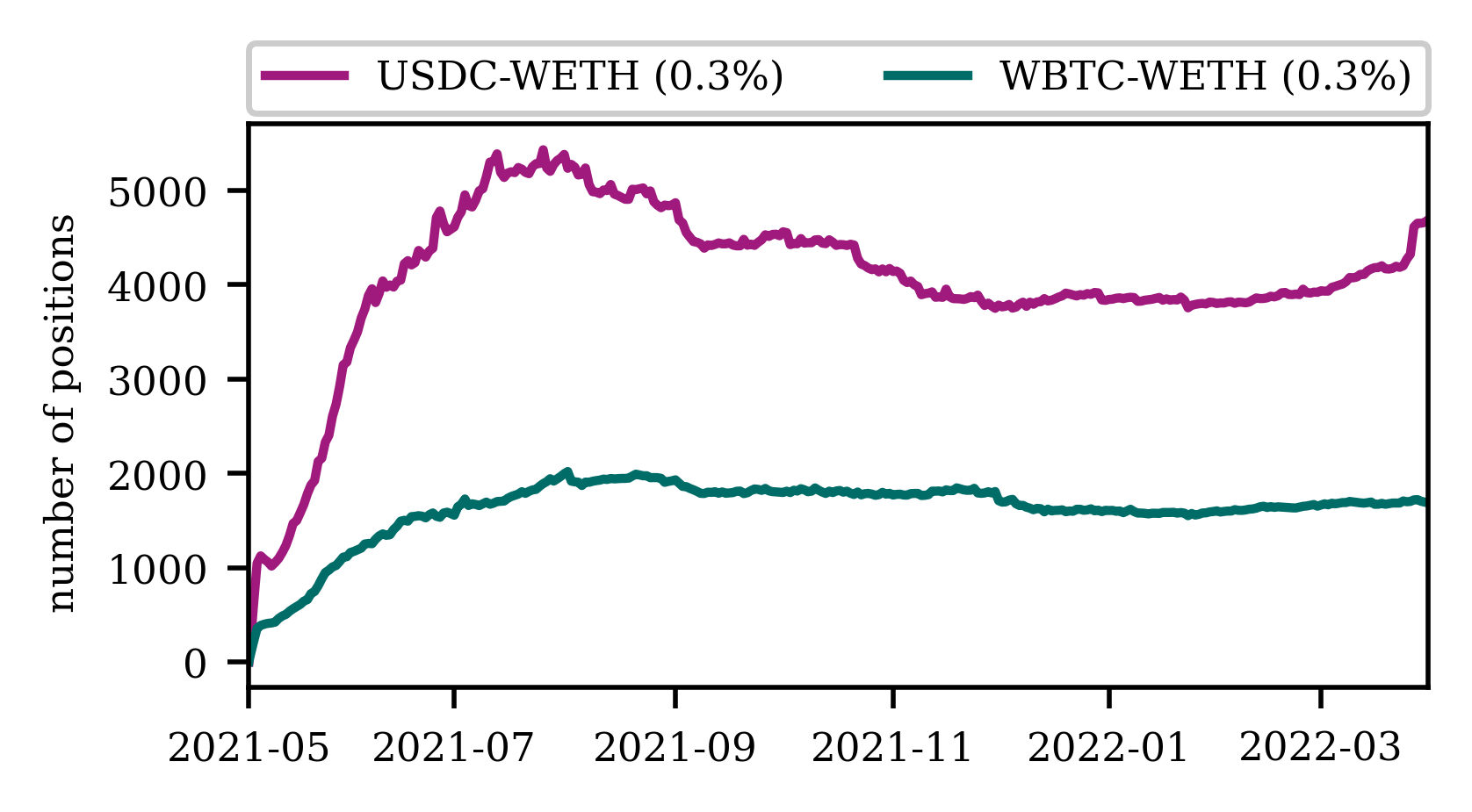}\vspace{-0pt}
\caption{USDC-WETH (f= 0.3\%) and WBTC-WETH (f= 0.3\%)} \label{fig:positionsETHBTC}
\end{subfigure}   
\begin{subfigure}[]{1\linewidth}
\includegraphics[scale=1,right]{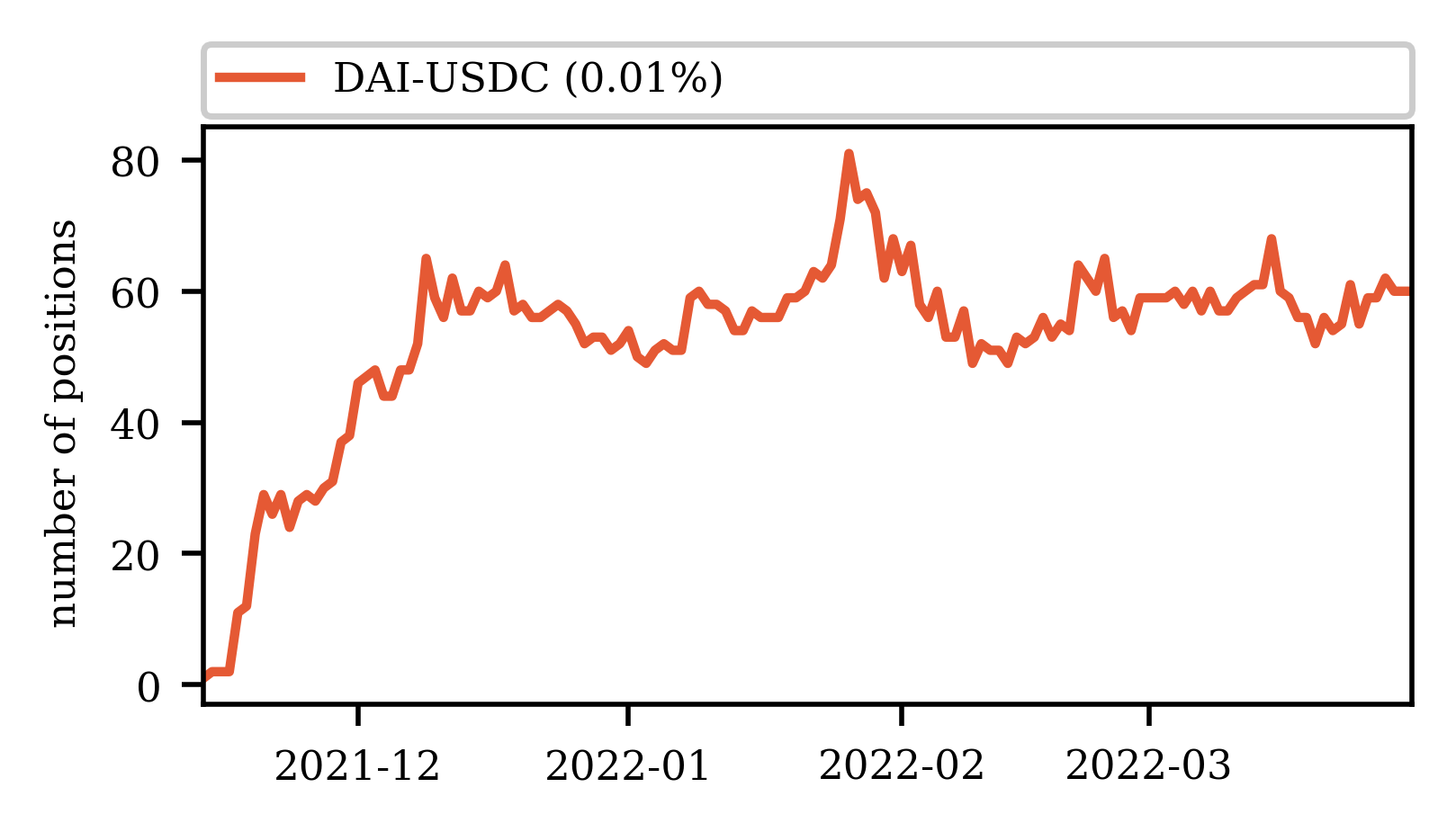}\vspace{-4pt}
\caption{DAI-USDC (f= 0.01\%)} \label{fig:positionsstable}
\end{subfigure} 
\caption{Number of active positions over time in three Uniswap V3 pools. Note the low number of active liquidity positions compared to the pool size for DAI-USDC.} \label{fig:positions}\vspace{-4pt}
\end{figure}

\begin{figure}[t]
\centering
\begin{subfigure}[]{1\linewidth}
\includegraphics[scale=1,right]{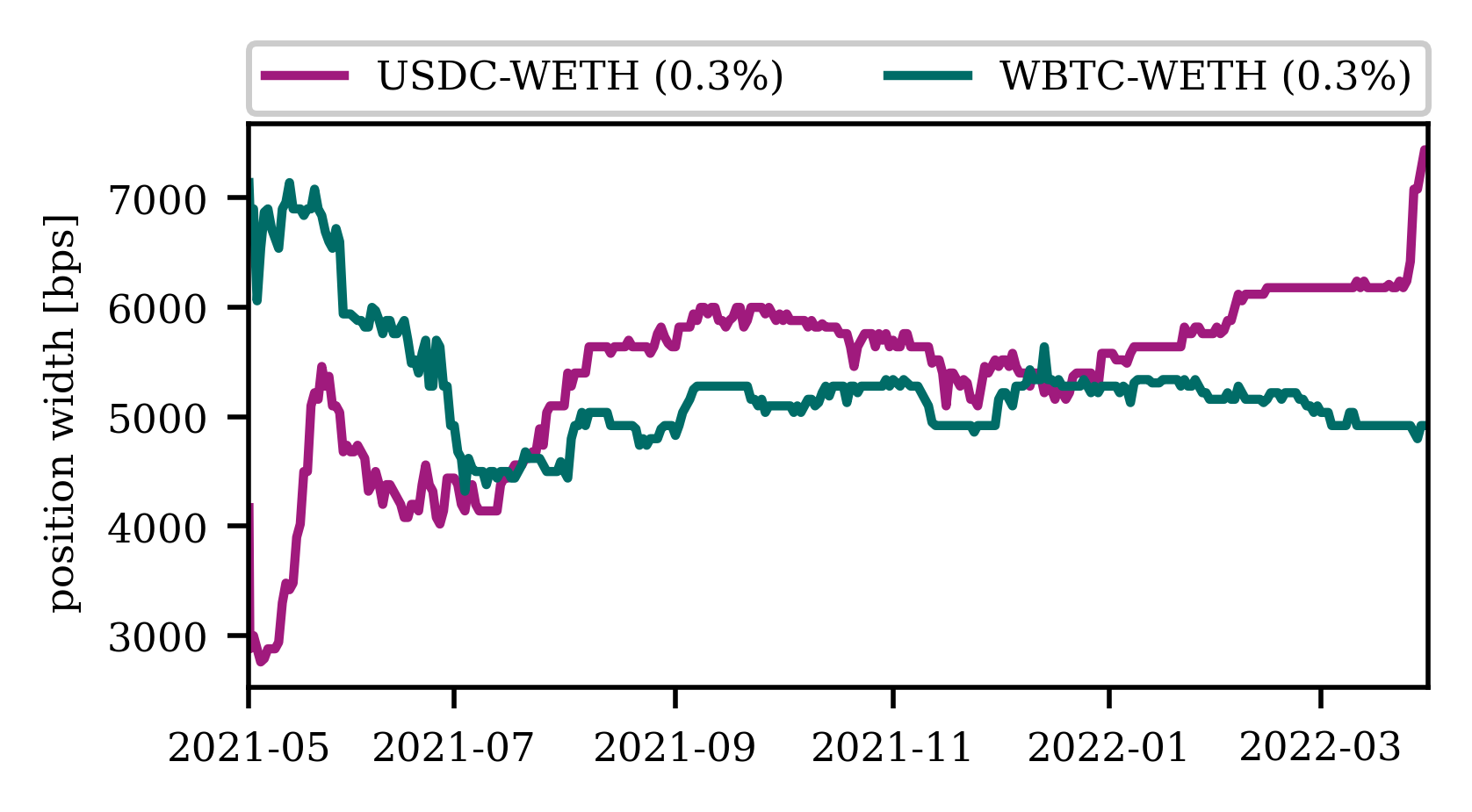}\vspace{-4pt}
\caption{USDC-WETH (f= 0.3\%) and WBTC-WETH (f= 0.3\%)} \label{fig:avgPositionSizeETHBTC}
\end{subfigure}   
\begin{subfigure}[]{1\linewidth}
\includegraphics[scale=1,right]{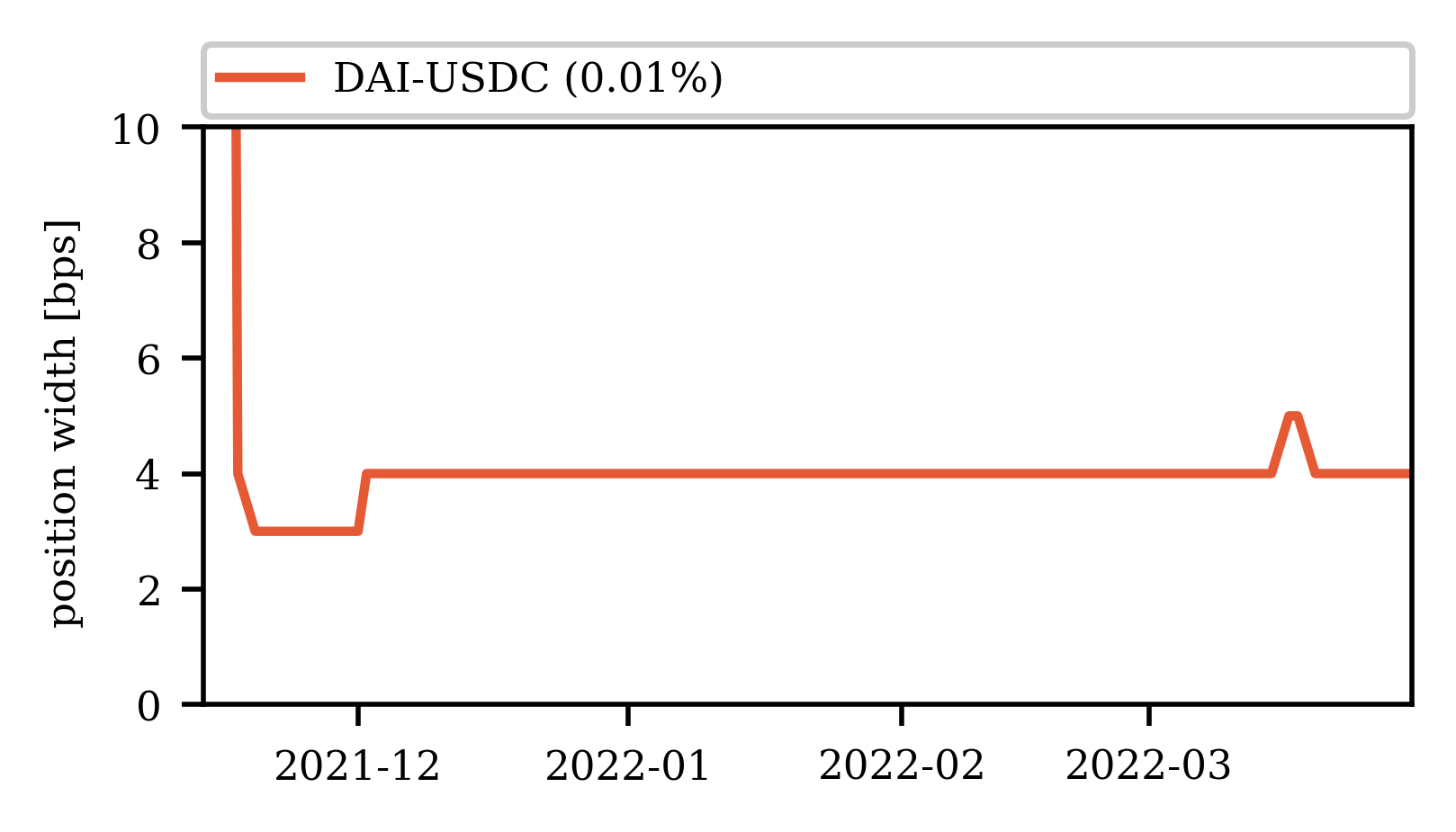}\vspace{-4pt}
\caption{DAI-USDC (f= 0.01\%)} \label{fig:avgPositionSizestable}
\end{subfigure} 
\caption{Median position width over time in three Uniswap V3 pools. } \label{fig:avgPositionSize}\vspace{0pt}
\end{figure}

Figure~\ref{fig:positionsize} shows both the median and mean position sizes over the pools' lifetime, while Figure~\ref{fig:positions} shows the number of active positions in each pool over time. Note that as the DAI-USDC (f= 0.01\%) was only created in late 2021 following the introduction of the new fee tier, the data set is significantly smaller (cf. Figure~\ref{fig:avgPositionSizestable}). When considering the position size statistics in the  USDC-WETH pool and WBTC-WETH pool (cf. Figure~\ref{fig:positionsizeETHBTC}), we notice that, apart from an initial growth phase, both the median (lighter lines) and the mean (darker lines) show little fluctuations over time. The median and mean position sizes are quite similar between the two pools. We also find that the number of active liquidity positions in the USDC-WETH pool is approximately double the number of active positions in the WBTC-WETH pool at all times. The number of liquidity positions in both pools' stabilizes at a couple of thousand after an initial growth period of around two months after the pool's creation (cf. Figure~\ref{fig:positionsETHBTC}). Finally, we observe that the mean is significantly (around ten times) larger than the median in both pools, indicating a highly unequal distribution of liquidity provider funds. When turning to the DAI-USDC pool (cf. Figure~\ref{fig:avgPositionSizestable}), we only observe this trend magnified. Until February 2022, the difference between the median and mean liquidity position size is also around a factor of ten but then increases to a factor of 100, indicating a pronounced discrepancy in the distribution of liquidity. We note that both the median and mean position sizes in the DAI-USDC pool are significantly larger than in the other two analyzed pools. Especially, the DAI-USDC pool thus appears to be in the hands of large liquidity providers: underlined by the extremely small number of active liquidity positions in the pool, as well as the presence of single liquidity positions worth more than US\$ 100'000'000. There are only around 60 active liquidity positions in the pool (cf. Figure~\ref{fig:positionsstable}), while the pool holds around US\$ 300'000'000 after its initial liquidity growth.

\begin{figure}[t]
\centering
\begin{subfigure}[]{1\linewidth}
\includegraphics[scale=1,right]{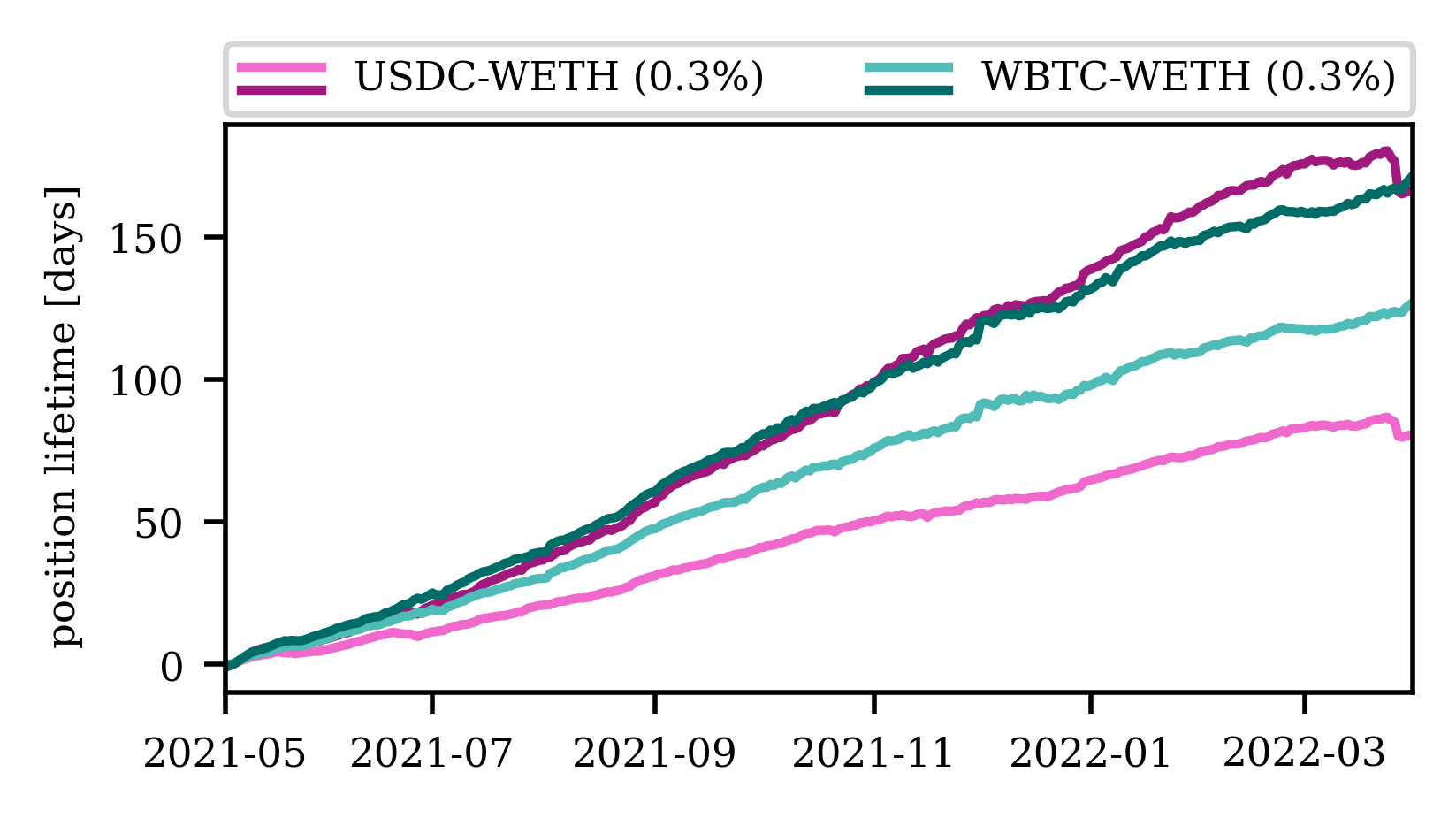}\vspace{-4pt}
\caption{USDC-WETH (f= 0.3\%) and WBTC-WETH (f= 0.3\%)} \label{fig:avgTimeInPositionETHBTC}
\end{subfigure}   
\begin{subfigure}[]{1\linewidth}
\includegraphics[scale=1,right]{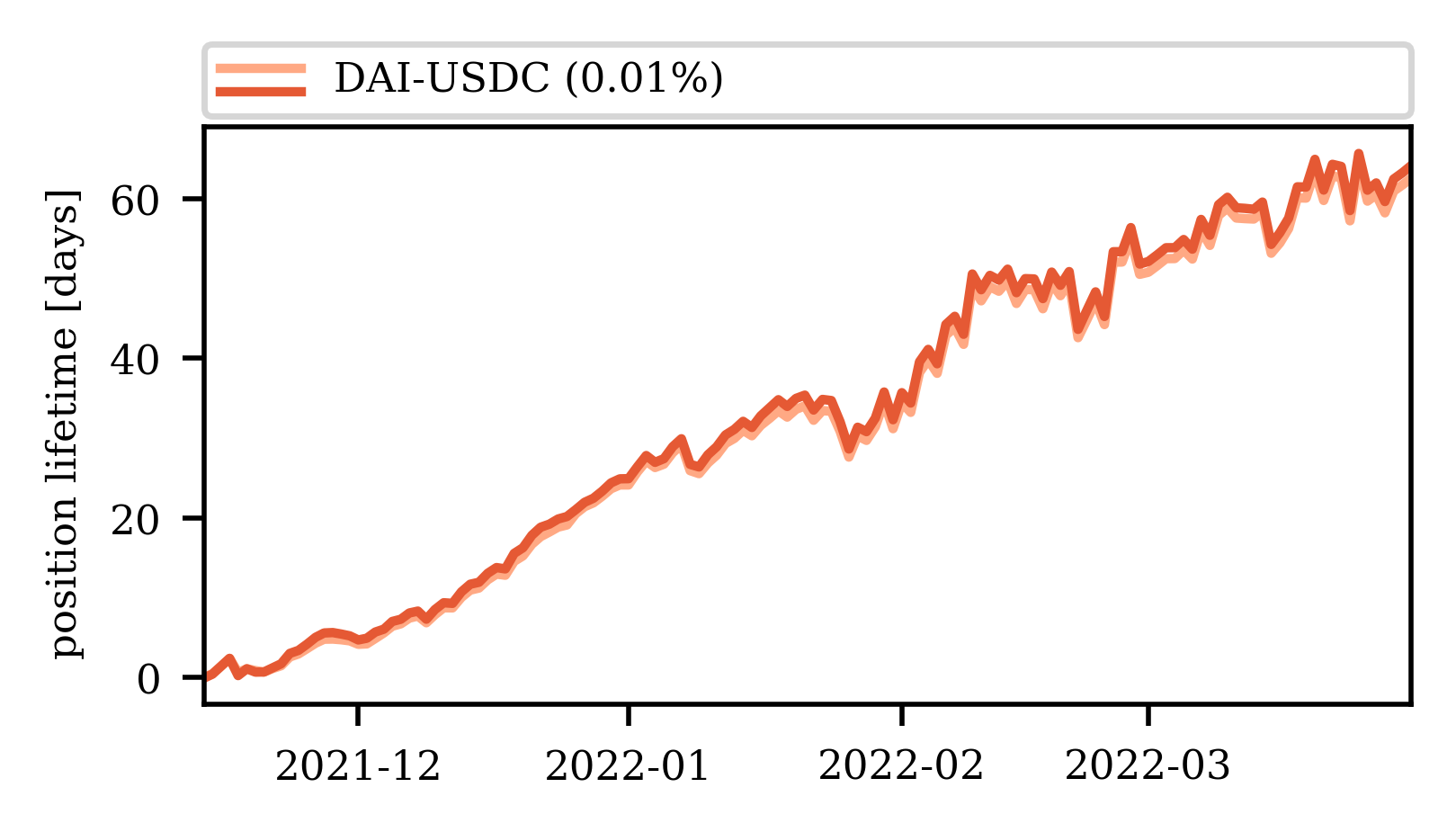}\vspace{-4pt}
\caption{DAI-USDC (f= 0.01\%)} \label{fig:avgTimeInPositionstable}
\end{subfigure} 
\caption{Mean time position was ITM (lighter lines) and total position lifetime (darker lines) over time in three Uniswap V3 pools.} \label{fig:avgTimeInPosition}\vspace{0pt}
\end{figure}

\begin{figure*}[t]
\centering

\begin{subfigure}[]{0.48\linewidth}
\includegraphics[scale=1,right]{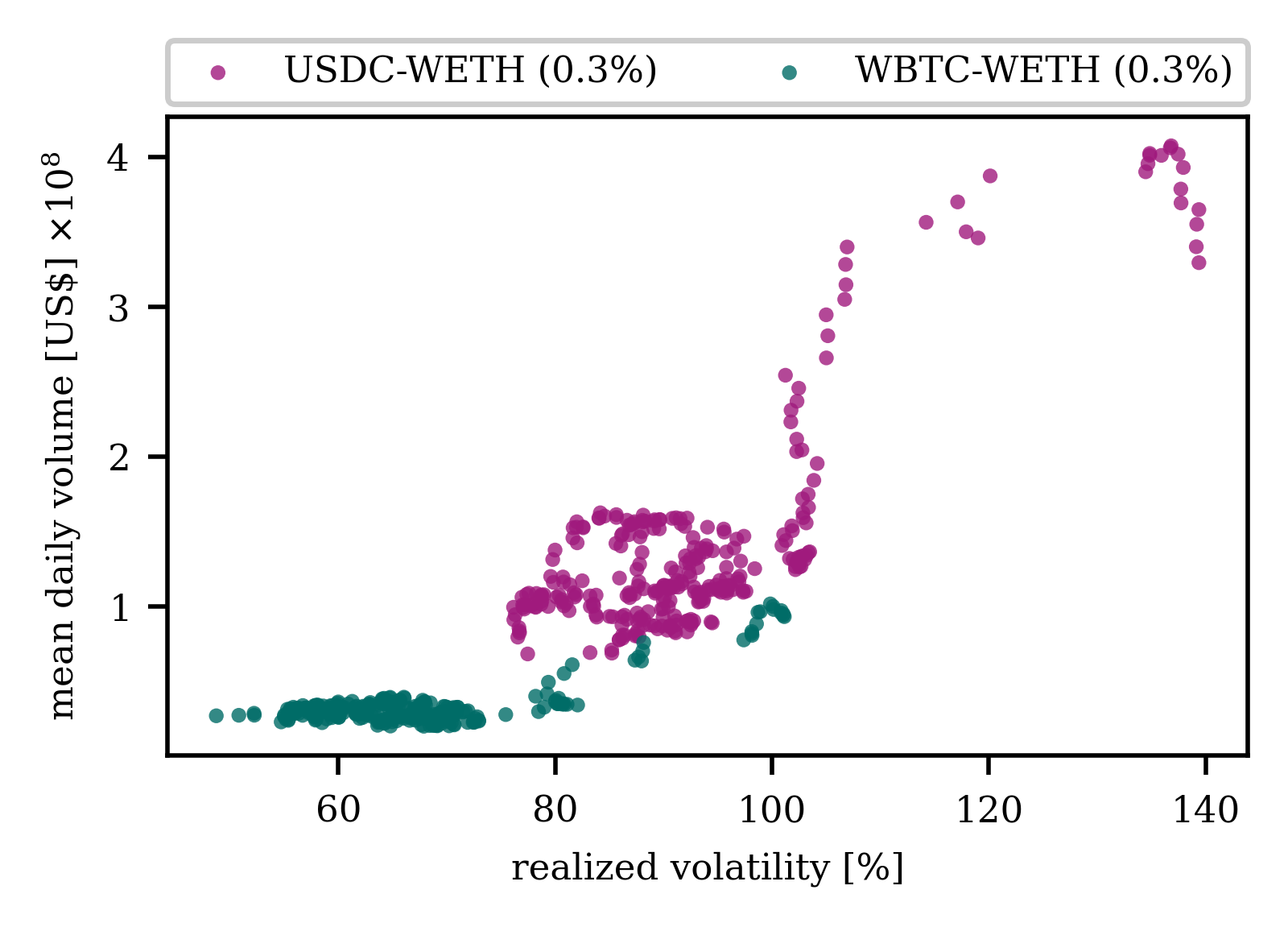}\vspace{-4pt}
\caption{USDC-WETH (f=0.3\%) and WBTC-WETH (f=0.3\%)} \label{fig:volvolETHBTC}\vspace{0pt}
\end{subfigure}
\hfill
\begin{subfigure}[]{0.48\linewidth}
\includegraphics[scale=1,right]{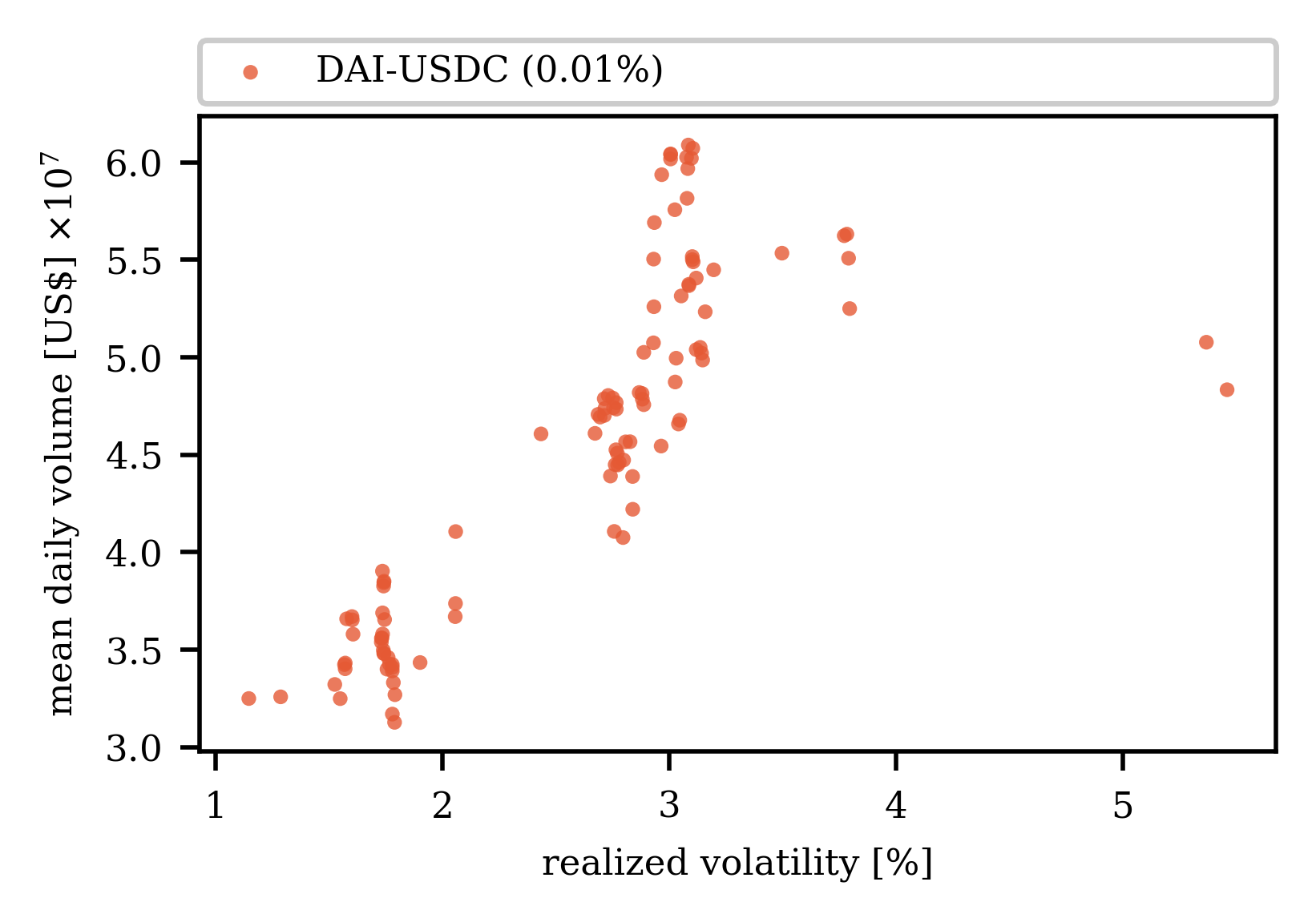}\vspace{-4pt}
\caption{DAI-USDC (f= 0.01\%)} \label{fig:volvolstable}\vspace{0pt}
\end{subfigure}   
\caption{Mean daily trading volume and realized volatility for each 30 day window in three Uniswap V3 pools.} \label{fig:volvol}\vspace{0pt}
\end{figure*}

In Figure~\ref{fig:avgTimeInPosition} we plot the mean position lifetime (darker lines), as well as the mean time spent by a position ITM (lighter lines) for the three analyzed pools. The mean position age increases linearly with the pool's lifetime at around half the rate for the three pools, indicating that a significant proportion of liquidity positions are active over a long period. For the two normal pairs (cf. Figure~\ref{fig:avgTimeInPositionETHBTC}), we observe a significant difference between a position age and the time the position was ITM, i.e., the position was active and earning fees. This difference is most pronounced for the USDC-WETH pool. In the pool, the mean of the time a position was active is only around half of the total position lifetime. Thus, on average, liquidity positions only earn fees during half the time. This difference is also present for the WBTC-WETH pool but less pronounced. In the WBTC-WETH pool, liquidity positions are, on average, active for more than two-thirds of their lifetime. We presume that this difference stems from the relative price between the two cryptocurrencies, Bitcoin and Ether, having a higher correlation with each other than with the US\$ (cf. Figure~\ref{fig:volvol}). The less volatile relative price makes it easier to determine the price range of a liquidity position and makes it less likely for the price to fall out of the liquidity position's price range. In the DAI-USDC pool, where both tokens are pegged to the US\$ and, thus, intended to have the same value at all times, the difference between the mean of a position lifetime and the mean of the time spent ITM is practically in-existent (cf. Figure~\ref{fig:avgTimeInPositionstable}).

In close relation, we consider the median position width, measured in bps, of liquidity positions in the three pools in Figure~\ref{fig:avgPositionSize}. Note that we consider the median as a sole position with an infinite price range that would have a too significant impact on the mean position width. We observe that for the stable pair (cf. Figure~\ref{fig:avgPositionSizestable}), the median position size is tiny with 4bps and almost constant soon after the pool's creation. The small price movements in the pool make it easy for liquidity providers to choose a capital-efficient, i.e., small, price range without the risk of the pool's price falling out of their price range. For the two normal pairs (cf. Figure~\ref{fig:avgPositionSizeETHBTC}), the median width of a liquidity position is significantly larger by a factor of around 1000. We further observe that while initially, the mean position size in the USDC-WETH pool was smaller than in the WBTC-WETH pool, the trend reverses over the pools' lifetimes. Thus, the market learns that it must select larger price ranges for liquidity positions in pools where the volatility of the relative price is larger. We further notice that, especially in the USDC-WETH pool, we observe an increase in the median position width, indicating that liquidity providers are becoming more familiar with Uniswap V3. Within a couple of months, liquidity providers as a whole appear to learn that they must select bigger price ranges if they want to hold their liquidity position for longer times, as we show in Section~\ref{sec:return}.

To conclude the general analysis of liquidity positions and pools on Uniswap V3, we consider both the pool's \emph{realized volatility} as well as its mean daily trading volume (Figure~\ref{fig:volvol}). The realized volatility measures the assets historic volatility. More specifically, the realized volatility $\sigma _r$ is given by: 
$$\sqrt{\frac{365}{T}\sum _{i=1}^T \ln{\frac{S_i}{S_{i-1}}}},$$
where $S_i$ is the pool's price, $T$ is the number of days over which $\sigma_r$ is measured, and the factor 365 scales to volatility to one year. Thus, for each 30 day window in our data set, we plot the realized volatility and the mean daily trading volume. We observe only small realized volatilities in the DAI-USDC (cf. Figure~\ref{fig:volvolstable}). In the USDC-WETH and WBTC-WETH pools, on the other hand, we observe a significant realized volatilities reaching 140\%. Notice that the realized volatility in the USDC-WETH pool is significantly larger than in the WBTC-WETH pool. While this finding might appear counter-intuitive initially, it stems from the prices of Bitcoin and Ether being correlated, and, therefore, leading to a less volatile relative price. In general, the higher the pool's price volatility, the higher the pool's acquired fees must be to compensate for the impermanent loss. In Figure~\ref{fig:volvol} we observe that for all three pools there is a correlation between the realized volatility and the mean daily trading volume over the same 30 day period. This correlation, exceeding 0.75 in each pool, is promising for liquidity providers. However, while we find this correlation within a pool, it is not reflected between pools. While liquidity providers must hope for higher trading volumes in more volatile liquidity pools, we find that the mean daily volume in the DAI-USDC pool is largely similar to that in the significantly more price volatile WBTC-WETH pool. Finally, it must be noted that while higher volatility may lead to higher trading volume and more fees collected, as seen in the previous section, the probability that the price drops out of the liquidity position also increases with volatility. Furthermore, volatility will also increase the probability of larger losses due to impermanent loss. Thus, high volatility is not per se in the interest of the liquidity provider.

\subsection{Performance Statistics of Liquidity Positions}\label{sec:performance}

We continue the analysis with an investigation of the performance of individual liquidity positions in six Uniswap V3 pools: USDC-WETH (f$\in$\{0.05\%,0.3\%\}), WBTC-WETH (f$\in$\{0.05\%,0.3\%\}) and DAI-USDC (f$\in$\{0.05\%,0.01\%\}). These pools include those with the highest volume as well as those with the highest liquidity on Uniswap V3~\cite{2021uniswap}. Further, we include all fee tiers of a given pair if the tier holds significant liquidity, allowing for additional comparisons. Throughout this section, we analyze all liquidity positions with a liquidity deposit in excess of US\$ 0.0001. Smaller positions can lead to erroneous returns due to the finite precision of ERC-20 tokens.

\begin{figure}[t]
\centering
\begin{subfigure}[]{1\linewidth}
\includegraphics[scale=1,right]{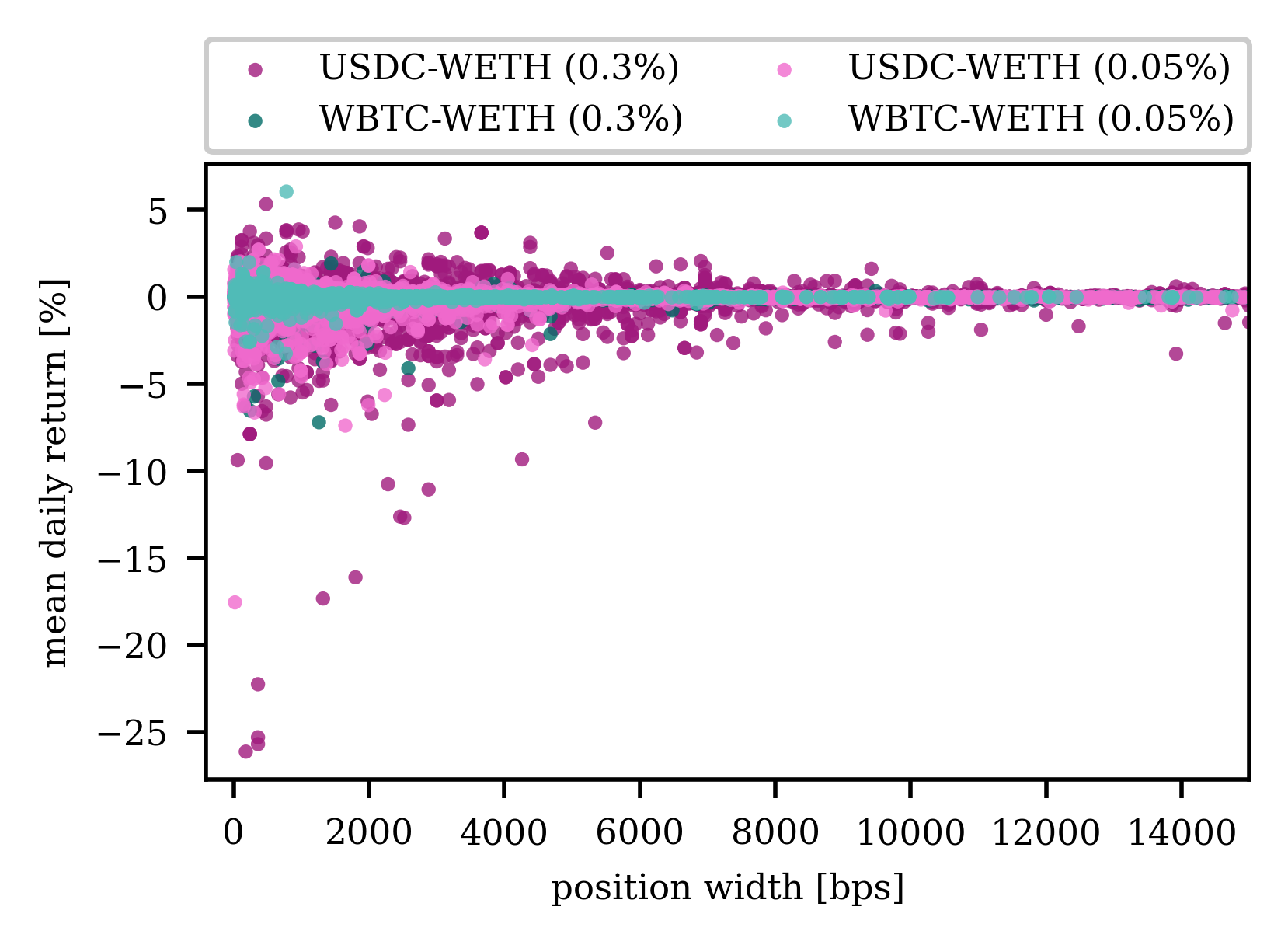}\vspace{-4pt}
\caption{USDC-WETH (f$\in$\{0.05\%,0.3\%\}) and WBTC-WETH (f$\in$\{0.05\%,0.3\%\})} \label{fig:returnpositionsizenormal}
\end{subfigure}
\begin{subfigure}[]{1\linewidth}
\includegraphics[scale=1,right]{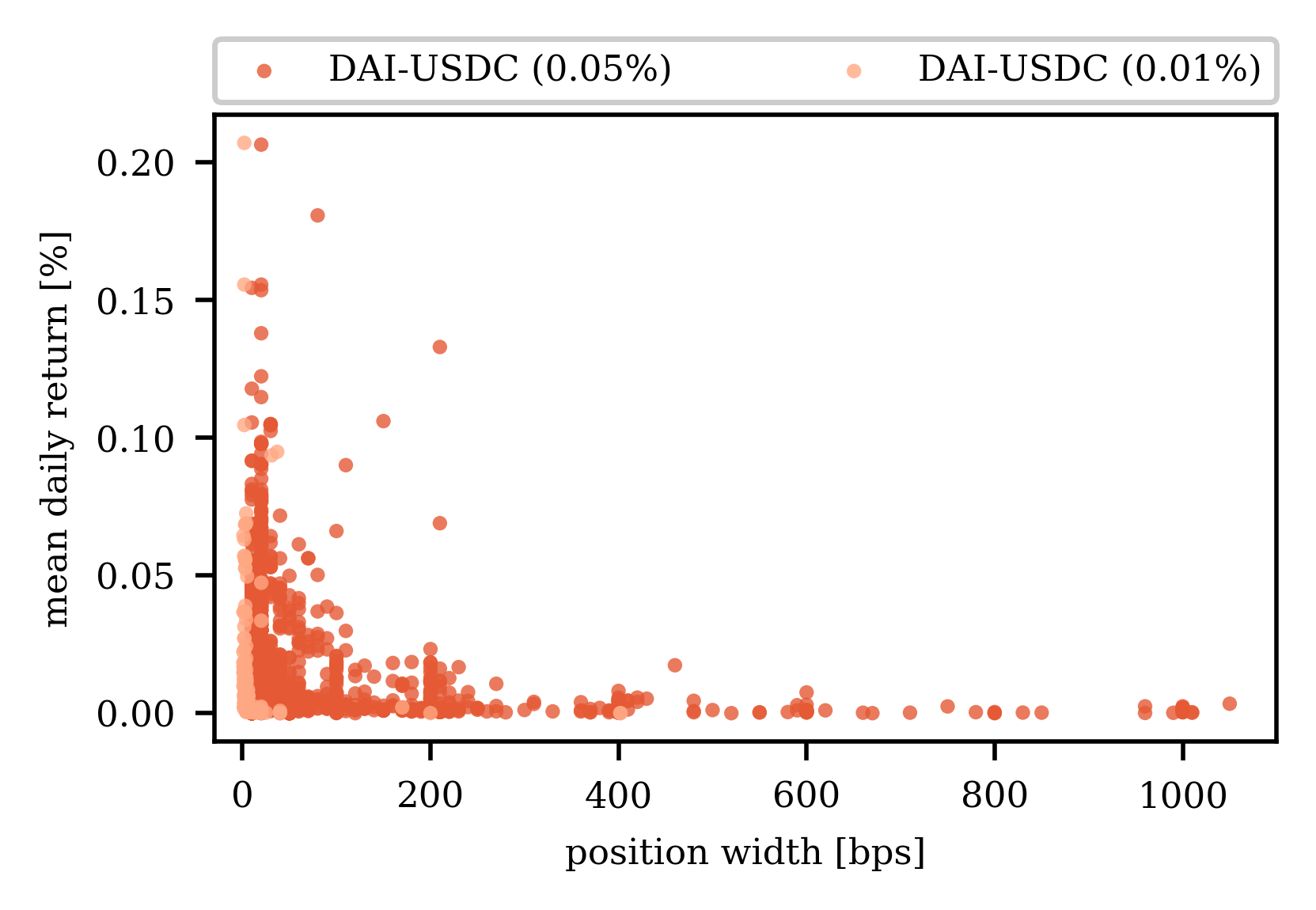}\vspace{-4pt}
\caption{DAI-USDC (f$\in$\{0.05\%,0.01\%\})} \label{fig:returnpositionsizestable}
\end{subfigure}   
\caption{Mean daily returns of individual liquidity positions, depending on the position width, in six Uniswap V3 pools. Observe the significant spread in daily returns for narrow positions. The mean of each of the data series in Figure~\ref{fig:returnpositionsizenormal} is negative, thus, on average, liquidity providers lose money in comparison to holding the assets and are hence not compensated for the additional risk of providing liquidity.} \label{fig:returnpositionsize}\vspace{0pt}
\end{figure}

In Figure~\ref{fig:returnpositionsize}, we plot the mean daily return as well as the width of the position's price range for all liquidity positions that were active for at least a day. Note that we calculate a position's daily return according to Equation~\ref{eq:return} at the end of each whole day during the position lifetime. A position's width represents a position's price range: the larger the width, the larger the price range. While vastly different patterns appear for normal pairs and stable pairs, we observe similar patterns within a category. Figure~\ref{fig:returnpositionsizenormal} shows the average daily return and position width of individual liquidity positions for the normal pairs. We notice that the magnitude of the mean daily returns can be significantly larger for small position widths than for large position widths, consistent with our prediction in Section~\ref{sec:return}. While the magnitude of the mean daily returns is significantly larger for small position widths, liquidity positions exhibit both positive and negative returns. The magnitude of the mean daily returns tends to be smaller in the slightly less price volatile WBTC-WETH pair pools, as well as in the respective pools with the smaller fee. In particular, we want to point out that there are a few liquidity positions with daily returns of around -20\% in the USDC-WETH pools. These positions were only active for a little longer than a day in mid-May of 2021. During this time, the price of Ether dropped by around 30\% on a single day~\cite{2022ethusdt}. For large price ranges the mean daily returns are very close to zero. Thus, liquidity providers can earn significant returns with small width liquidity positions in normal pairs, but at the same time, the risks are higher. Choosing larger liquidity positions minimizes risks, but the mean daily returns are unlikely to be significant. 

For the stable pair a wildly different pattern appears (cf. Figure~\ref{fig:returnpositionsizestable}z. While the magnitude of the daily returns also decreases with the position width, they are never significantly negative. Due to the negligible price volatility in both stable pair pools, there is no impermanent loss, the driving factor of negative returns. Additionally, while the mean daily returns are more significant for liquidity positions with small price ranges, they only reach about 0.3\% in the most extreme cases and are, thus, significantly less than for normal pairs.

\begin{figure}[t]
\centering
\begin{subfigure}[]{1\linewidth}
\includegraphics[scale=1,right]{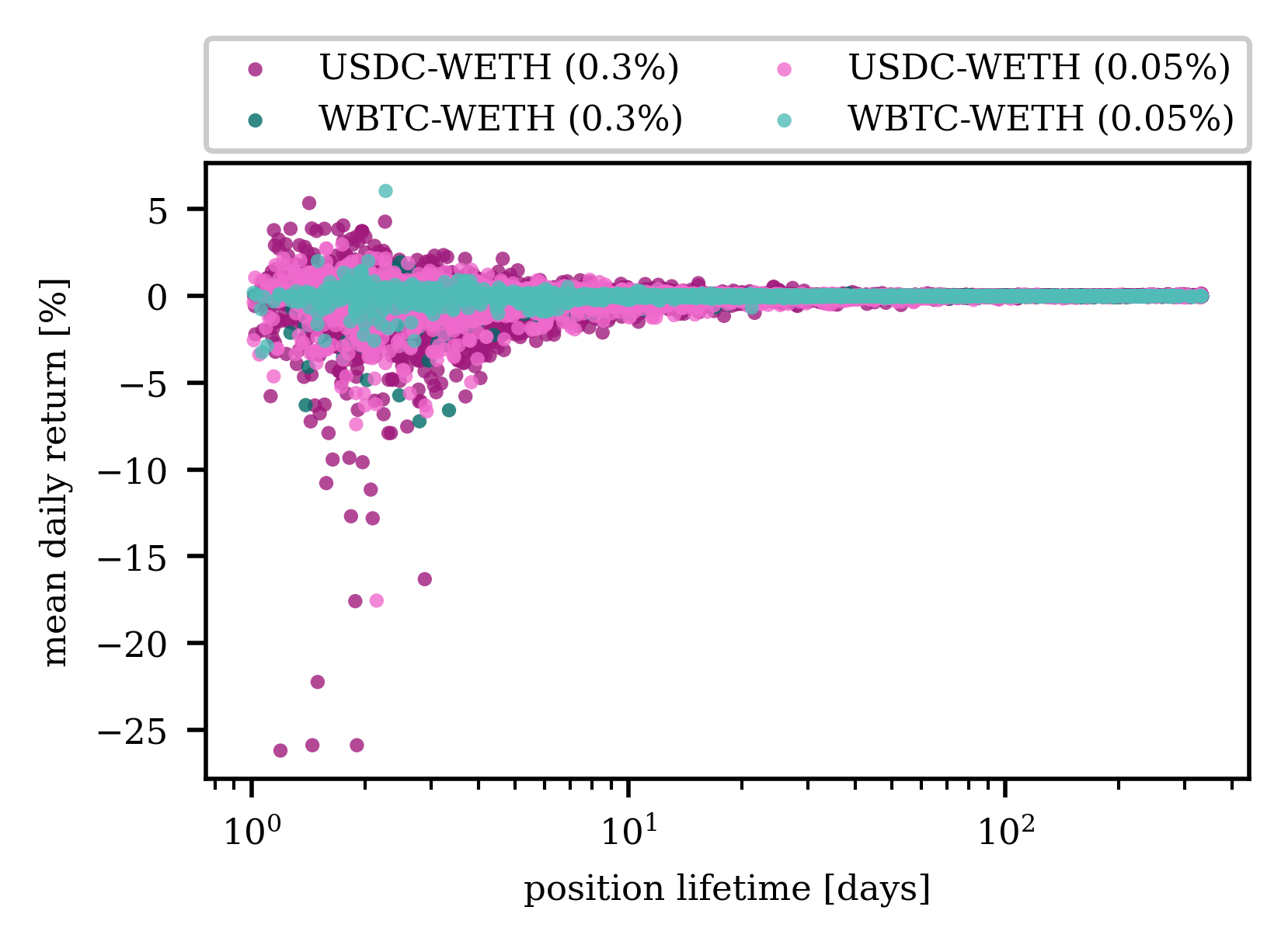}\vspace{-4pt}
\caption{USDC-WETH (f$\in$\{0.05\%,0.3\%\}) and WBTC-WETH (f$\in$\{0.05\%,0.3\%\})} \label{fig:lifetimeETHBTC}
\end{subfigure}
\begin{subfigure}[]{1\linewidth}
\includegraphics[scale=1,right]{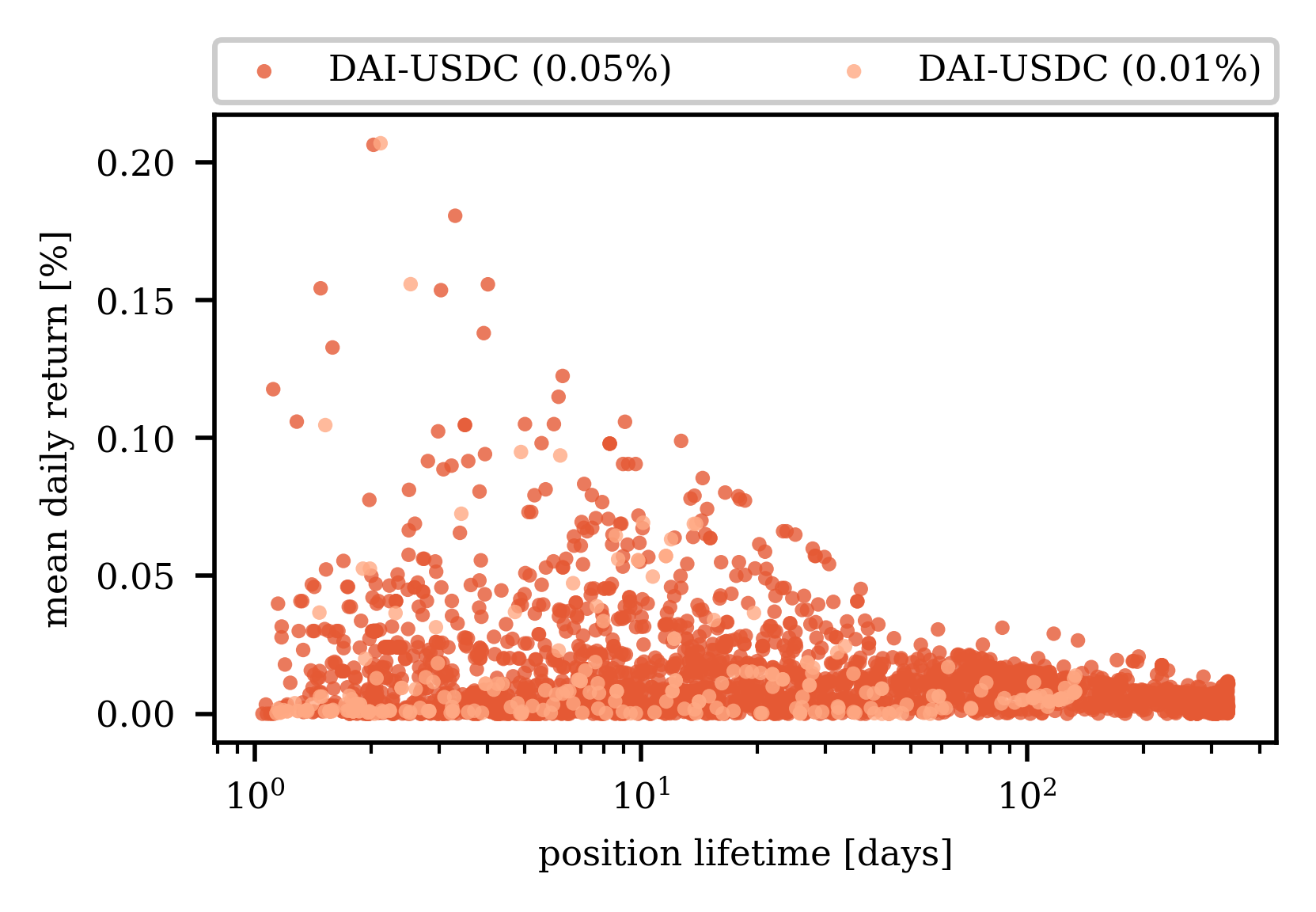}\vspace{-4pt}
\caption{DAI-USDC (f$\in$\{0.05\%,0.01\%\})} \label{fig:lifetimestable}
\end{subfigure}   
\caption{Mean daily returns of individual liquidity positions as a function of the position lifetime in six Uniswap V3 pools.} \label{fig:lifetime}\vspace{0pt}
\end{figure}

The difference between normal and stable pairs is as apparent when plotting the mean daily returns of individual liquidity positions against the lifetime of a position (cf. Figure~\ref{fig:lifetime}). While the magnitude of the daily return tends to decrease with the lifetime of a position, this trend is more apparent in the four normal pair pools (cf. Figure~\ref{fig:lifetimeETHBTC}). Only liquidity position with a short lifetime tend to experience stark positive and negative daily returns. Thus, garnering significant profits as a liquidity provider requires active management, indicating that providing liquidity in Uniswap V3 is a game reserved to professional traders. In Figure~\ref{fig:lifetimestable}, we observe that more extreme values for the mean daily return values are only present for liquidity position's with a shorter lifetime. These outliers stem from variations in the pool's daily volume. In general most liquidity positions exhibit similar daily returns independent of their lifetime. As liquidity provider returns are mainly influenced by the pool's volume and available liquidity, daily fluctuations of returns are less significant. 

\begin{figure}[t]
\centering
\begin{subfigure}[]{1\linewidth}
\includegraphics[scale=1,right]{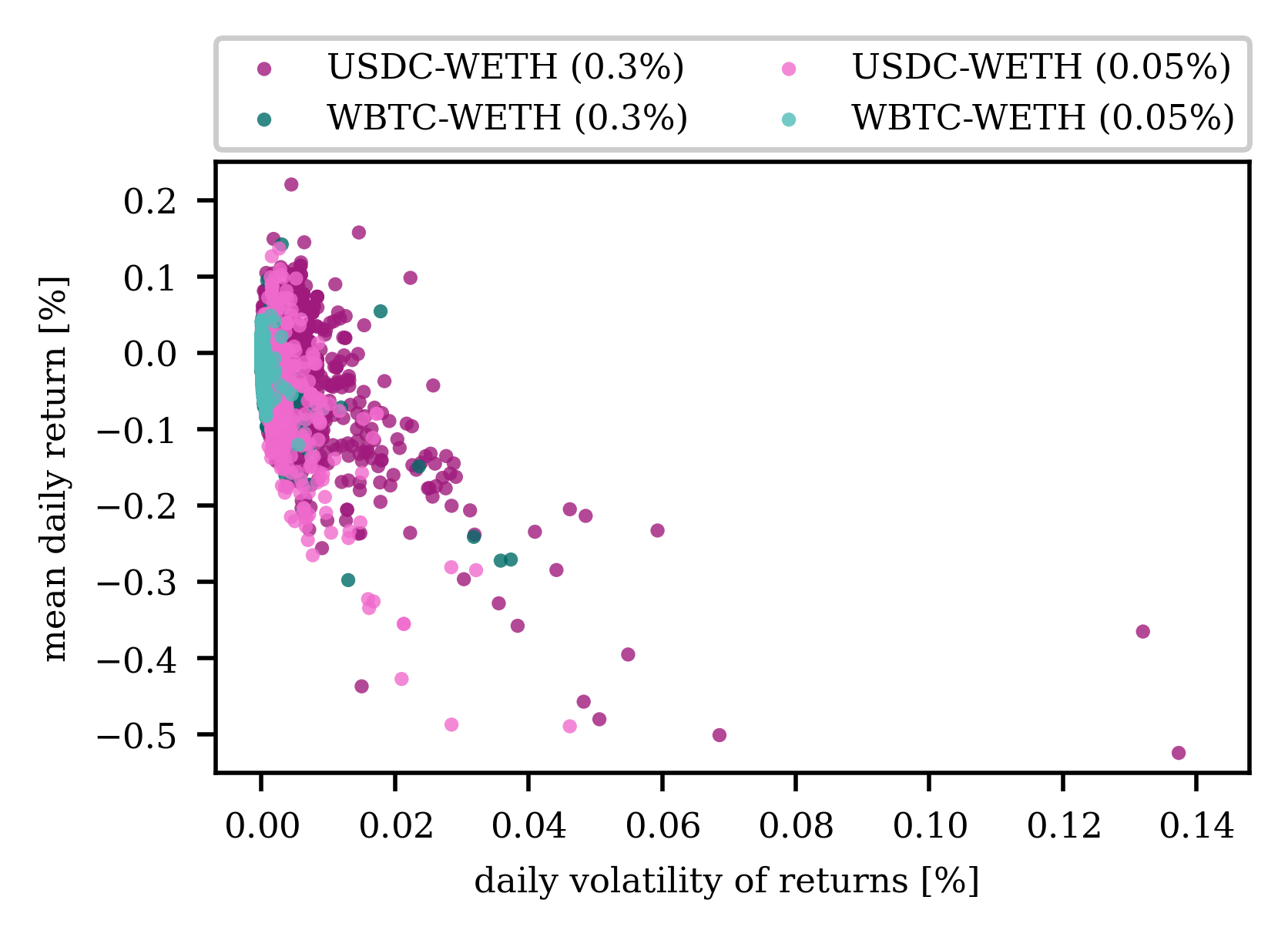}\vspace{-4pt}
\caption{USDC-WETH (f$\in$\{0.05\%,0.3\%\}) and WBTC-WETH (f$\in$\{0.05\%,0.3\%\})} \label{fig:volETHBTC}
\end{subfigure}
\begin{subfigure}[]{1\linewidth}
\includegraphics[scale=1,right]{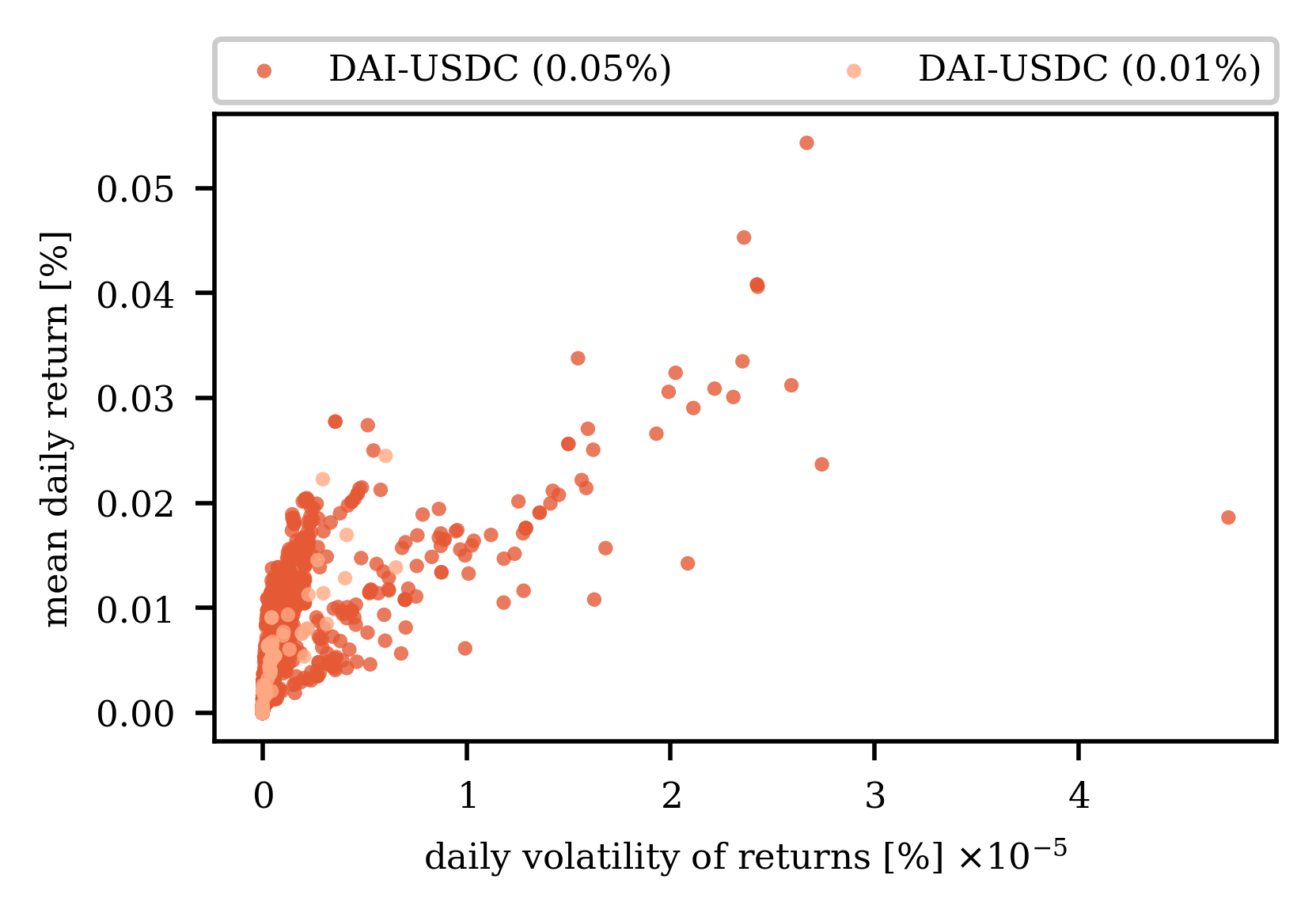}\vspace{-4pt}
\caption{DAI-USDC (f$\in$\{0.05\%,0.01\%\})} \label{fig:volstable}
\end{subfigure}   
\caption{Mean daily returns and volatility of daily returns of individual liquidity positions for six Uniswap V3 pools.}\label{fig:vol}\vspace{0pt}
\end{figure}

The negligible volatility of the daily returns received by liquidity positions in the two stable pair pools is further highlighted in the risks and return analysis of liquidity positions in Figure~\ref{fig:vol}. We plot the daily mean returns against the volatility of the returns for each liquidity position in Figure~\ref{fig:vol} in order to analyze the risks and returns of liquidity positions. Note that we only include positions whose lifetime exceeds 30 days to allow for more representative calculations of return volatility. Thus, positions with short lifetimes, that generally experience the most extreme daily returns (cf. Figure~\ref{fig:lifetime}), are not in the data set. Generally speaking, higher volatility in the returns of an investment suggests greater risks. We find that in the two stable pair pools (cf. Figure~\ref{fig:volstable}), the daily volatility of the daily returns is incredibly small. Additionally, all liquidity positions with lifetimes exceeding 30 days exhibit mean daily returns ranging from 0\% to 0.04\%. Thus, while liquidity position's on Uniswap V3 experience little to no risks of losing money, the returns are generally small. For the four normal pair pools (cf. Figure~\ref{fig:volETHBTC}), we, as expected, observe more significant mean daily returns also for positions with lifetimes exceeding a month. However, this comes at the cost of a higher volatility of the daily returns, suggesting greater risks. Note that this is in line with our expectation, as the risks presented to liquidity providers stem from the impermanent loss, which is driven by price fluctuations between the pair's two assets. We also observe that liquidity positions in the USDC-WETH experience more extreme volatility in their daily returns than in the WBTC-WETH pools. This observation is in line with both the higher fluctuation in price between US\$ and Ether, as well as with the higher fluctuations in volume (cf. Figure~\ref{fig:volvol}). We further notice that for normal pairs liquidity positions that have the highest volatility have the poorest performance. The opposite is true for stable pairs where the positions with the highest volatility experience the most significant daily returns. 

\begin{figure}[t]
\centering
\begin{subfigure}[]{1\linewidth}
\includegraphics[scale=1,right]{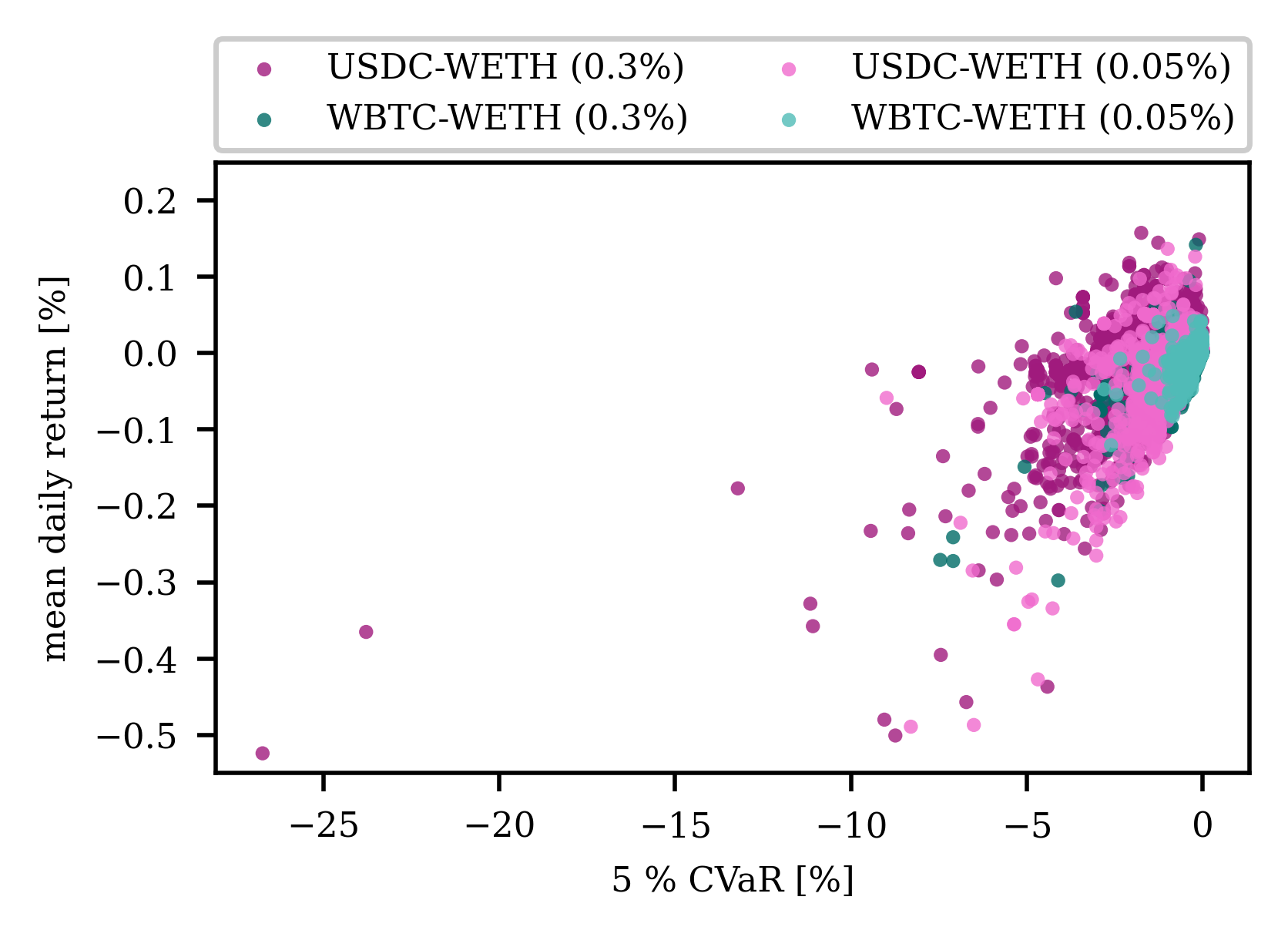}\vspace{-4pt}
\caption{USDC-WETH (f$\in$\{0.05\%,0.3\%\}) and WBTC-WETH (f$\in$\{0.05\%,0.3\%\})} \label{fig:cvarBTCETH}
\end{subfigure}
\begin{subfigure}[]{1\linewidth}
\includegraphics[scale=1,right]{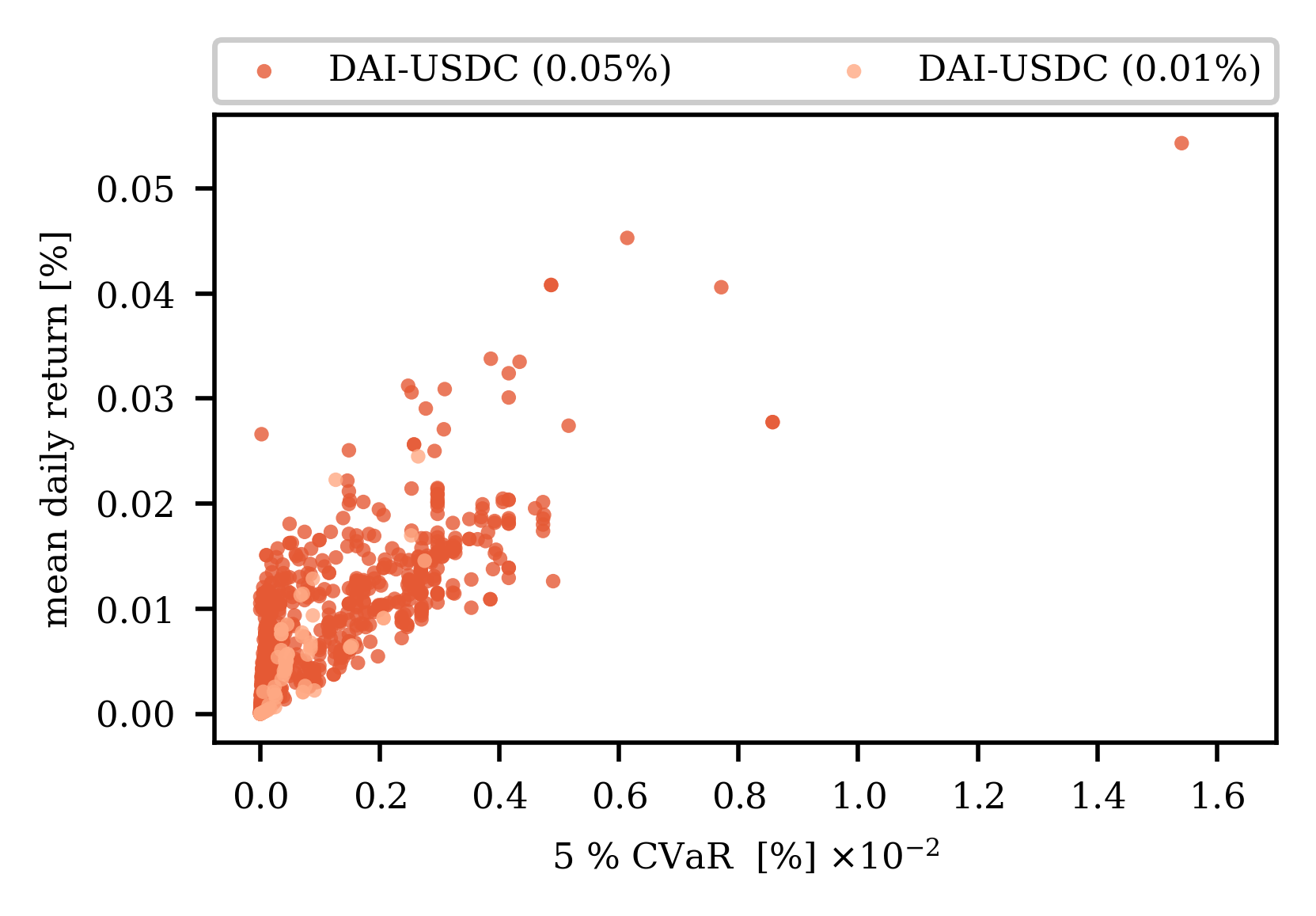}\vspace{-4pt}
\caption{DAI-USDC (f$\in$\{0.05\%,0.01\%\})} \label{fig:cvarstable}
\end{subfigure}  
\caption{Mean daily returns of individual liquidity positions and 5\% CVaR of daily returns for six Uniswap V3 pools.} \label{fig:cvar}\vspace{0pt}
\end{figure}

To further study the risks associated with providing liquidity, we also consider the \emph{conditional value at risk (CVaR)} of the analyzed liquidity positions (cf. Figure~\ref{fig:cvar}). The 5\% CVaR represents the expectation value of the return of an investment in the 5\% worst cases~\cite{rockafellar2000optimization}. The CVaR is one of the most frequently used risk measures, as it is not influenced by higher than average returns and best reflects the tail-risk behavior of investments. It is therefore sometimes referred to as the average worst case loss. We observe that the CVaR is positive for the vast majority of liquidity positions in the two stable pair pools (cf. Figure~\ref{fig:cvarstable}). Thus, even in the 5\% worst cases, the expected daily return of the liquidity positions is still positive, again suggesting a very moderate financial risk for liquidity providers in these pools. For the four normal pair pools, a different picture paints itself (cf. Figure~\ref{fig:cvarBTCETH}). Individual positions experience CVaRs worse than -10\%, exemplifying the risk related to providing liquidity in non-stable pools. While individual positions experience these extreme CVaRs, the 5\% CVaR of most positions is better than -5\%, but these higher risks experienced by liquidity providers do not appear to be rewarded with high returns. Less than 30\% of the liquidity positions in the four pools are rewarded for the added risks they shoulder in comparison to providing liquidity in the stable pools, signaling that achieving high returns is not a simple undertaking as a Uniswap V3 liquidity provider.

We note here that while we measured low historical volatilities in the stablecoin pools in our analysis, this must not hold true in the future. The low volatility in stablecoin pools relies on the continued confidence investors place in the respective stablecoins. The price turbulences of UST in May 2022~\cite{2022terraust} only highlight that providing liquidity should not be viewed as a passive investment -- even in stablecoin pools. Both the market design of Uniswap V3 and the necessary decision-making from liquidity providers may pose a challenge to the protocol. Sudden and unexpected price changes, such as seen in May 2022 for UST or USDT, can cause the price to move such that little or no liquidity depth is available at the market price, causing trading on Unsiwap V3 to cease~\cite{2021usdcudt,2021usdcusdt}. 



\section{Conclusion}
Providing liquidity in traditional finance markets is generally an investment form reserved to professional traders and institutions. The decentralized nature of the blockchain, on the other hand, allows for many individual liquidity providers to join together to facilitate trustless cryptocurrency exchanges on the blockchain while earning fees. Previous works have shown that providing liquidity on DEXes utilizing the original CPMM design can be a profitable investment, which is accessible to retail traders and only requires a few simple considerations from their side.

In contrast, our work shows that obtaining high returns as a liquidity provider on Uniswap V3 is a highly complicated undertaking requiring active management and a good know-how. The introduction of Uniswap V3 has thus turned liquidity providing into a playing field for sophisticated investors where retail traders must be wary to avoid risking significant losses. Retail traders, unwilling to risk these losses and unable to perform the resource intensive active management, should therefore restrict themselves to simple strategies that yield only small returns. 
\balance

\bibliographystyle{ACM-Reference-Format}
\bibliography{references}



\end{document}